\documentclass[12pt,preprint]{aastex}
\usepackage[numberedappendix]{emulateapj5} \usepackage{times,mathptm}
\usepackage{danonecolfloat}
\slugcomment{Draft: \today}
\shorttitle{Halo Model Constraints from Angular Biases}
\shortauthors{Scranton}

\def\rs{r_{\rm S}}
\def\rhobar{\bar{\rho}}
\def\nbar{\bar{n}}
\def\nbarr{\bar{n}_{\rm R}}
\def\nbarb{\bar{n}_{\rm B}}
\def\nuprime{\nu^{\prime}}
\def\numgal{\langle N \rangle}
\def\numgalr{\langle N \rangle_{\rm R}}
\def\numgalb{\langle N \rangle_{\rm B}}
\def\secgal{\langle N(N-1) \rangle}
\def\secgalr{\langle N(N-1)\rangle_{\rm R}}
\def\secgalb{\langle N(N-1)\rangle_{\rm B}}
\def\fourgal{\langle N \rangle^{(4)}}
\def\Nx{N_{\rm X}}

\def\Mr{M_{\rm R}}
\def\Mc{M_{\rm R0}}
\def\Mb{M_{\rm B}}
\def\Mbs{M_{\rm Bs}}
\def\gammar{\gamma_{\rm R}}
\def\gammab{\gamma_{\rm B}}
\def\rhos{\rho_{\rm S}}
\def\rhor{\rho_{\rm SR}}
\def\rhob{\rho_{\rm SB}}
\def\alphar{\alpha_{\rm R}}
\def\alphab{\alpha_{\rm B}}
\def\yr{y_{\rm R}}
\def\yb{y_{\rm B}}

\def\fr{f_{\rm R}}
\def\fb{f_{\rm B}}
\def\brb{b^2_{\rm RB}(\theta)}
\def\brg{b^2_{\rm RG}(\theta)}
\def\bbg{b^2_{\rm BG}(\theta)}

\def\Plin{P_{\rm LIN}(k)}
\def\Pxxz{P_{\rm XX}(k,\chi)}
\def\Pxx{P_{\rm XX}(k)}
\def\Pgg{P_{\rm GG}(k)}
\def\Prr{P_{\rm RR}(k)}
\def\Pbb{P_{\rm BB}(k)}

\def\Txxxx{T_{\rm 4X}(k_1,k_2,k_3,k_4)}
\def\PKxx{{\cal P}_{\rm XX}(K)}

\def\TKxxxx{{\cal T}_{\rm 4X}(K_1,K_2,K_3,K_4)}
\def\TbarKxx{\bar{{\cal T}}_{\rm XX}(K_1,K_2)} 

\def\Phhgg{P^{hh}_{\rm GG}(k)}
\def\PPgg{P^{P}_{\rm GG}(k)}
\def\Phhrr{P^{hh}_{\rm RR}(k)}
\def\PPrr{P^{P}_{\rm RR}(k)}
\def\Phhbb{P^{hh}_{\rm BB}(k)}
\def\PPbb{P^{P}_{\rm BB}(k)}

\def\Fx{F_{\rm X}(\chi)}
\def\Fr{F_{\rm R}(\chi)}
\def\Fb{F_{\rm B}(\chi)}
\def\wxx{w_{\rm XX}(\theta)}
\def\wgg{w_{\rm GG}(\theta)}
\def\wrr{w_{\rm RR}(\theta)}
\def\wbb{w_{\rm BB}(\theta)}
\def\bb{\hat{b}^2(\theta)}
\def\gprime{g^\prime}
\def\rprime{r^\prime}
\def\iprime{i^\prime}

\begin{document}
\twocolumn[

\title{Testing the Halo Model Against the SDSS Photometric Survey}


\author{Ryan Scranton}
\affil{Department of Astronomy and Astrophysics, University of Chicago, 
Chicago, IL 60637 USA}
\affil{NASA/Fermilab Astrophysics Center, P.O. Box 500, Batavia, IL 60510 USA}
\email{scranton@oddjob.uchicago.edu}

\begin{abstract}
We present halo model predictions for the expected angular clustering and 
associated errors from the completed Sloan Digital Sky Survey (SDSS) 
photometric galaxy sample.  These results are used to constrain halo 
model parameters under the assumption of a fixed $\Lambda$CDM cosmology using 
standard Fisher matrix techniques.  Given the ability of the five-color SDSS 
photometry to separate galaxies into sub-populations by intrinsic color, we 
also use extensions of the standard halo model formalism to calculate the 
expected clustering of red and blue galaxy sub-populations as a further test 
of the galaxy evolution included in the semi-analytic methods for populating 
dark matter halos with galaxies.  The extremely small sample variance and 
Poisson errors from the completed SDSS survey should result in very impressive 
constraints ($\sim 1-10\%$) on the halo model parameters for a simple 
magnitude-limited sample and should provide an extremely useful check on the 
behavior of current and future N-body simulations and semi-analytic 
techniques.  We also show that similar constraints are possible using a narrow 
selection function, as would be possible using photometric redshifts, without 
making linear assumptions regarding the evolution of the underlying power 
spectra.  In both cases, we explore the effects of uncertainty in the 
selection function on the resulting constraints and the degeneracies between 
various combinations of parameters.

\end{abstract}



]


\section{Introduction}

In the standard picture of structure formation, initial perturbations in the 
dark matter density collapse into halos (White \& Rees, 1978; White \& Frenk, 
1991) in a hierarchical manner, starting at small mass scales and moving to 
larger masses over time.  Simulations of this process have found that these 
halos have a self-similar shapes (Navarro, Frenk \& White, 1996, NFW, 
hereafter; Moore et al., 1998), provided that one allows for some difference
in central density for halos as a function of mass (Navarro, Frenk \& White, 
1997).  Following this distribution of dark matter, we expect the baryonic 
matter to fall into these halos, cool and eventually form galaxies.  This 
process can be simulated as well and the results can likewise be modeled
by relatively simple semi-analytic methods (Kauffmann et al., 1999; 
Somerville \& Primack, 1999; Benson et al., 2000).  In addition to determining
where galaxies form within a dark matter halo, these methods can also give 
estimates of the morphology, color, and star formation rates for galaxies, 
under certain assumptions.  

The development of these prescriptions for the distribution of dark matter and
galaxies has allowed for the calculation both the real-space two-point 
functions (Sheth \& Jain, 1997; Jing et al., 1998; Peacock \& Smith 2000) and 
power spectra (Seljak, 2000; Scoccimarro et al., 2000; Ma \& Fry, 2000), 
generating predictions for galaxy clustering statistics in both the linear 
and nonlinear regimes which are both physically well-motivated and quite 
simple.  These treatments can be further generalized to accommodate different 
galaxy sub-population clustering (Scranton, 2002; S02, hereafter), allowing 
for more detailed calculations.

On the data side, the next generation of galaxy surveys, in particular the 
Sloan Digital Sky Survey (SDSS; York et al. 2000; Gunn et al. 1998; 
Fukugita et al. 1996) have recently begun producing the anticipated large, 
rich galaxy catalogs.  The initial galaxy clustering measurements (Zehavi et 
al., 2001; Connolly et al., 2002; Scranton et al., 2002; Gaztanaga 2001) have 
demonstrated not only the remarkable quality of data possible with a fully 
digital large area survey but also the enormous statistical power which will 
be available from such a combination of large area and depth of redshift.  At 
the same time, the constraints on cosmological parameters ($\Omega_M$, 
$\Omega_\Lambda$, $\sigma_8$, etc.) have been improved by recent measurements 
of the CMB (cf. Pryke et al. (2001)) and large scale structure measurements 
(cf. Tegmark et al. (2001)) to the extent that we can reasonably consider the 
details of galaxy clustering and evolution in the context of a fixed cosmology.

The simultaneous development of these powerful tools for exploring 
galaxy clustering and evolution leads one to consider the possibilities for 
testing the predictions from the halo model against the future prospects of the
data.  Toward this end, we present calculations of the expected constraints on 
the halo model parameters from measurements of the angular clustering of 
galaxies for the completed SDSS survey.  Since we wish to test not only the 
ability of the halo model to predict general galaxy clustering but also the 
modeling of galaxy evolution, we extend the general calculation to include 
clustering of the expected red and blue sub-populations.  The exploration of 
sub-population clustering has been done to a certain extent in the SDSS 
redshift survey (Zehavi et al., 2001).  However, many of the halo model 
parameters we will be investigating affect the respective power spectra on 
small scales ($k > \sim 1 h {\rm Mpc}^{-1}$), a region where spectroscopic 
surveys are plagued by redshift-space distortions and observational 
complications (e.g. the collisions between spectroscopic fibers in the SDSS 
survey).  Although these effects can be mitigated to a certain extent by 
projecting the clustering along the line of sight, one can achieve similar 
effects by considering the angular clustering in the photometric catalog.  
This approach also offers the benefit of a much deeper look ($z \sim 0.3$) at 
the clustering for all galaxies than the SDSS main galaxy spectroscopic sample 
($z \sim 0.1$) is capable of delivering.

From the standpoint of the halo model, the importance of constraining the halo
model parameters with observations is two-fold.  The various components of the 
halo model (mass function, concentration, etc.) have all been parameterized 
in various ways and fit to the results of simulations which try to replicate 
the underlying physics as close as is feasible.  Our approach mirrors those 
efforts, replacing the simulation input with expected measurements on the sky. 
From the standpoint of the halo model as a theoretical construct, this sort of 
constraint will provide an invaluable check on whether the parameterizations 
coming out of simulations reasonably correspond to the data on the sky.  If 
so, then we have a powerful tool for analytic calculations and future insight
into the development of large scale structure and galaxy evolution.  If not, 
we can isolate those parts of the formalism have failed (determining that the 
disagreement between simulation outputs and the observations is due to a 
failure in the assignment of galaxies to halos rather than the halo mass 
function, for instance).  This feeds into the second aspect of the 
constraints: improving future simulations.  The simulations carry greater 
information about the underlying physics than is present in the 
parameterizations that feed into the halo model calculations.  Replacing 
simulation constraints on the halo model parameters with ones taken from the 
data should indicate which aspects of the physics that are included in the 
simulations are necessary to generate the observed clustering and evolution 
and what might be ignored.

In \S\ref{sec:fiducial}, we review the basics of the halo model and the 
augmentations necessary to calculate sub-population power spectra, establishing
the fiducial set of model parameters we will use for all our calculations.  
With this laid out, \S\ref{sec:selection_function} discusses the two types of
selection functions we will use to project the three-dimensional galaxy 
clustering onto the sky.  \S\ref{sec:wtheta} briefly discusses the formalism
for calculating the angular correlations as well as the corresponding 
covariance matrices and the additional information needed to account for 
uncertainties in the selection function.  These angular correlations are 
converted into angular biases in \S\ref{sec:bias}, with their own attendant 
covariance matrix.  These are then fed into the Fisher matrix formalism 
described in \S\ref{sec:fisher} to produce constraints on each of the 
parameters.  Finally, \S\ref{sec:results} discusses the results of these 
calculations, the expected errors and degeneracies on each of the halo model 
parameters.

\section{The Fiducial Model} \label{sec:fiducial}

For the purpose of calculating the power spectra that will feed into our 
angular correlations, there are two broad classes of parameters we will 
consider (we leave the details of how the parameters combine to produce power 
spectra to Appendix~\ref{sec:power}).  First, we have the general halo model 
parameters which describe the overall dark matter halo profile and biasing.  
These parameters have been measured from a number of N-body simulations and 
we will adopt conventional values.  

The second set of parameters describes the halo occupation density: the 
abundance and distribution of galaxies (and galaxy sub-populations) in the 
halos.  As mentioned previously, a variety of semi-analytic techniques have 
been applied to N-body simulations.  For the purpose of our calculations, we 
will use the outputs of the GIF simulations (Kauffmann et al., 1999) which 
have been generated using the SDSS magnitudes.  These particular 
implementations of the GIF methodology have not been the subject of extensive 
inquiry in the literature, giving us more flexibility to explore the 
parameterization possibilities.  Further, given the relatively light 
computational load of calculating eventual constraints on these 
parameters, we will allow ourselves a generous parameter space, larger, in 
fact, than could easily be constrained by an actual measurement of the angular 
clustering for red and blue galaxies given simple computational methods.  
However, by performing this seemingly over-zealous calculation, we can 
determine which of the parameters are well-constrained by the angular 
correlations and which might be fixed without significantly affecting the 
errors on the other more sensitive parameters.  

\subsection{Dark Matter Halo Parameters}

Beginning with the general halo parameters, the fundamental unit of the 
halo model is the halo profile.  This can be parameterized along the lines
of the profile derived by NFW,
\begin{equation}
\rho(r) = \frac{\rhos}
{ \left ( r/ \rs \right )^{-\alpha} \left ( 1 + r/ \rs \right )^{(3+\alpha)}}
\label{eq:NFW}
\end{equation}
where $\rs$ is the universal scale radius and 
$\rhos = 2^{3+\alpha} \rho(\rs)$.  We can replace $\rs$ by a concentration, 
$c \equiv r_{\rm v}/\rs$, where $r_{\rm v}$ is the virial radius.  This radius 
is defined as the radius encompassing a region within which the fractional 
overdensity of the halo $\Delta_{\rm V}$ (Eke, et al., 1996) scales as
\begin{equation}
\Delta_{\rm V}(z) = 18 \pi^{2} (\Omega_M(z)^{-0.55}),
\end{equation}
where $\Omega_M(z)$ is the matter density relative to the critical density
for a given redshift,
\begin{equation}
\Omega_M(z) = \left [1 + \frac{1 - \Omega_M}{(1+z)^3 \Omega_M} \right]^{-1},
\end{equation}
$\Omega_M$ is the matter density today relative to the critical density, and 
we have chosen a $\Lambda$CDM cosmology where $\Omega_M + \Omega_\Lambda = 1$. 
The concentration is a weak function of halo mass ($c\equiv c_0(M/M_*)^\beta$, 
where $c_0 \sim {\cal O}(10)$ and $\beta \sim -{\cal O}(10^{-1})$).  The 
traditional NFW profile gives $\alpha = -1$, while the Moore profile has 
$\alpha = -3/2$.  We will use $\alpha = -1.3$ for the calculations in this 
paper, but the general results are largely insensitive to the choice of 
$\alpha$.  Bullock et al. (2001) gives $c_0 = 9$ for a pure NFW 
profile; using Peacock \& Smith's relation, $c_0 \approx 4.5$ for a Moore 
profile.  Since we are using an intermediate value of $\alpha$, we choose 
$c_0 = 6$ and $\beta = -0.15$ for all the calculations in this paper.  In 
order to generate power spectra at a variety of redshifts, we also scale the
concentration as $c_0 \sim (1 + z)^{-1}$ for a given redshift $z$, as in 
Bullock et al. (2001).

Once we know how the mass in a halo is distributed, we need to know how many
halos of a given mass we expect to find, i.e. the halo mass function.  
Traditionally, this mass function is expressed in terms of a function $f(\nu)$,
\begin{equation}
\frac{dn}{dM} dM = \frac{\rhobar}{M} f(\nu) d\nu,
\end{equation}
where $\nu$ relates the minimum spherical over-density that has collapsed at a
given redshift ($\delta_c$) and the rms spherical fluctuations containing 
mass $M$ ($\sigma(M)$) as
\begin{equation}
\nu \equiv \left (\frac{\delta_c}{\sigma(M)} \right )^2,
\label{eq:nu_def}
\end{equation}
This can be generalized for an arbitrary redshift by taking the forms from 
Navarro, Frenk and White (1997),
\begin{equation}
\delta_c(z) = \frac{3}{20} (12\pi)^{2/3}(\Omega_M(z)^{0.0055}) 
\end{equation}
and scaling $\sigma(M)$ as $\sigma(M,z) \equiv \sigma(M,0) D(z)$ 
where $D(z)$ is the linear growth factor for a given redshift $z$ normalized 
to unity at $z = 0$.  We define $M_*$ as the mass corresponding to $\nu = 1$.  
The functional form for $f(\nu)$ is traditionally given by the Press-Schechter 
function (1974).  This form tends to over-predict the number of halos below 
$M_*$, so we use the form found from simulations by Sheth and Tormen (1997),
\begin{equation}
\nu f(\nu) \sim (1 + {\nuprime}^{-p}){\nuprime}^{1/2} e^{-{\nuprime}/2},
\label{eq:dndM}
\end{equation}
where $\nuprime = a\nu$, $a = 0.707$ and $p = 0.3$.  This gives us a total of
five general halo model parameters we might constrain with angular 
measurements: $\alpha$, $c_0$, $\beta$, $a$ and $p$.

\begin{figure}[b]
\begin{center}
\epsfxsize=240pt \epsfbox{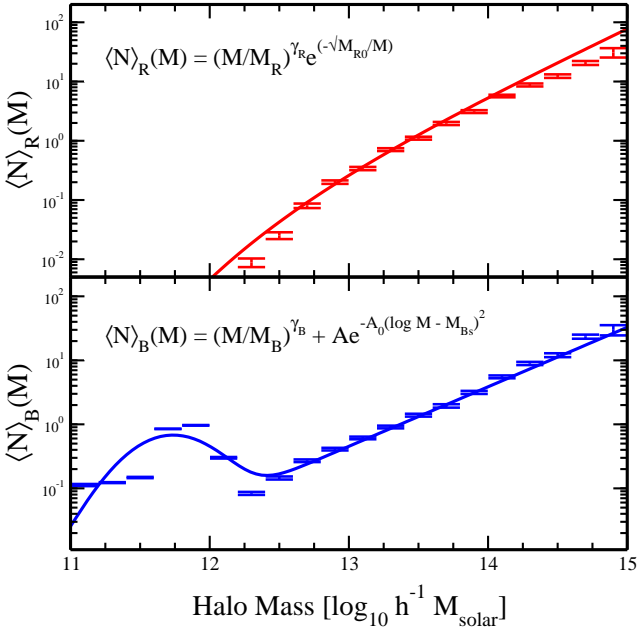}
\caption{$\numgal(M)$ for red and blue galaxies as measured at $z=0.27$ in the 
GIF simulations.  The solid lines indicate the values of $\numgalr$ and
$\numgalb$ using the forms in Equations~\ref{eq:numgal_red} and 
\ref{eq:numgal_blue} and parameter values in Table~\ref{tab:fiducial}.  The 
errorbars in both cases represent the Poisson errors in each mass bin.}
\label{fig:ngal}
\end{center}
\end{figure}

\subsection{Galaxy Parameters}\label{sec:split}

As described in Appendix~\ref{sec:power}, to calculate the galaxy power 
spectrum, we need to know the mean number of galaxies for a given halo mass
($\numgal (M)$) and the second moment of the galaxy distribution 
($\secgal (M)$).  Following the example in S02, we will generate different 
power spectra for red and blue galaxies by using different $\numgal$ 
relations for each of the sub-populations as well as changing the distribution
of the galaxies within each halo.  Before settling on the specific values of
the respective parameters, we need to establish some necessary formalism.

Figure~\ref{fig:ngal} shows the measurements of $\numgal(M)$ taken from the
GIF simulations for red and blue galaxies.  As in S02, we can parameterize the
$\numgal (M)$ relations as modified power laws.   In the case of red galaxies, 
the power law is cut-off by a lower mass limit 
($\Mc$),
\begin{equation}
\numgalr(M) = \left ( \frac{M}{\Mr} \right )^{\gammar} 
e^{ - \left (\Mc/M \right)^{1/2}}. 
\label{eq:numgal_red}
\end{equation}
For the blue galaxies, the modification is a bit more involved, including a 
Gaussian term to account for low-mass halos with a single blue galaxy.  Thus,
our $\numgalb (M)$ relation is 
\begin{equation}
\numgalb(M) = \left (\frac{M}{\Mb} \right )^{\gammab}
+ A e^{-A_0(\log(M) - \Mbs)^2} \label{eq:numgal_blue}.
\end{equation}
As mentioned above, we must also adopt a form for the second moment of the 
distributions, $\secgal (M)$.  In S02, we followed the results found by 
Scoccimarro et al. (2001) which related $\secgal$ to $\numgal$
as $\secgal (M) = \alpha_M^2(M) \numgal^2(M)$, where 
\begin{equation}
\alpha_M (M) = \left\{ \begin{array}{ll} 1 & M > 10^{13}h^{-1}M_{\sun} \\
\log \left (\sqrt{M/10^{11}h^{-1}M_{\sun}} \right) &  M < 10^{13}h^{-1}M_{\sun}
\end{array} \right .\label{eq:alpha_M}
\end{equation}
This gives the galaxies the Poisson distribution at large mass and sub-Poisson
distribution observed in simulations.  Since the mass dependence of $\alpha_M$ 
can significantly affect the small scale power, we would like to include this 
effect in our parameter constraints.  However, to make the form easier to 
incorporate into our later Fisher matrix formalism, we use a sigmoid 
function, 
\begin{equation}
\alpha_M (M) = \left [ 1 + e^{(\log(M) - M_\alpha)/\sigma_\alpha} 
\right ]^{-1}, \label{eq:secgal}
\end{equation}
where $M_\alpha$ and $\sigma_\alpha$ are chosen to match the behavior in the 
GIF simulations (fitting to match the behavior in Equation~\ref{eq:alpha_M}
results in values of 12 and 0.45 for $M_\alpha$ and $\sigma_\alpha$, 
respectively). 

\begin{table}[t]
\begin{center}
\caption{Fiducial model parameters}
\label{tab:fiducial}
\begin{tabular}{@{\extracolsep{\fill}}c|c|c}
Class & Parameter & Value \\
\hline
{\it Dark Matter Halo} & $\alpha$ & -1.3 \\  
& $c_0$ & 6 \\
& $\beta$ & 0.15 \\
& $a$ & 0.707 \\
& $p$ & 0.3 \\ \hline
{\it Galaxy HOD} & $\mu$ & 50.6 \\  
& $\eta$ & 3 \\  
& $\Mr$ & $1.8 \times 10^{13}h^{-1}M_{\sun}$ \\  
& $\gammar$ & 1.1 \\
& $\Mc$ & $4.9 \times 10^{12}h^{-1}M_{\sun}$ \\ 
& $\Mb$ & $2.34 \times 10^{13}h^{-1}M_{\sun}$ \\ 
& $\gammab$ & 0.93 \\  
& $A$ & 0.65 \\  
& $A_0$ & 6.6 \\  
& $\Mbs$ & 11.73  \\ 
& $M_\alpha$ & 12.1 \\
& $\sigma_\alpha$ & 0.27 \\ \hline 
\end{tabular}
\end{center}
NOTES.---%
Galaxy Halo Occupation Density (HOD) parameters taken from GIF simulations at 
$z = 0.27$.  Red galaxies taken to have $g^\prime - i^\prime > 0.85$ in rest
frame colors.

\end{table}

While Equations~\ref{eq:numgal_red} and \ref{eq:numgal_blue} are sufficient
to describe the expected number of red and blue galaxies in a halo of given 
mass, they do not suffice to match the observation that red and blue galaxies
have different radial distributions within a given halo, with red galaxies 
tending to populate the halo center and blue galaxies the outer regions. To 
generate different distributions for our red and blue galaxies within each
halo, we follow the method outlined in S02, using profiles of the form given 
in Equation~\ref{eq:NFW} as the distribution functions for the red and blue 
galaxies within each halo.  In principle, one could construct distribution 
functions for each of the galaxy sub-populations from observations of their
relative abundances at a number of radii.  However, by assuming a form for the
distributions, the power-law indices and relative normalizations of these 
distributions are effectively determined by considering two parameters: the 
ratio of blue to red galaxies at large halo radii ($\eta$) and the inverse 
ratio at small radii ($\mu$).  Since these ratios relate to the number of 
galaxies, rather than the mass assigned to the galaxies, we have to transform 
these quantities through our $\numgal$ relations from 
Equations~\ref{eq:numgal_red} and \ref{eq:numgal_blue} in order to maintain 
mass conservation within the halo.  Thus, we define $\eta^\prime$ and 
$\mu^\prime$ as
\begin{eqnarray}
\label{eq:eta_prime} 
\eta^\prime &=& \left[ \eta - A e^{-A_0(\log(\Mr)-\Mbs)^2} \right] ^{1/\gammab}
\frac{\Mb}{\Mr} \\ 
\mu^\prime &=& 
\left [ \mu \left (1 + Ae^{-A_0(\log(\Mb)-\Mbs)^2} \right) \right ]^{1/\gammar}
\frac{\Mr}{\Mb}. \nonumber
\end{eqnarray}
For large radii, the profiles will both scale as $\rho \sim r^{-3}$, so 
$\eta^\prime$ sets the relative normalization directly:
\begin{eqnarray}
\label{eq:rho_prime} 
\rhor &=& \frac{1}{\eta^\prime + 1} \rhos \\
\rhob &=& \frac{\eta^\prime}{\eta^\prime + 1} \rhos, \nonumber
\end{eqnarray}
The choice of radius ($r_i$) for the measurement of $\mu$ is somewhat 
arbitrary, so we follow S02 in setting $r_i c/r_{\rm v}$ = 0.1, (where $c$ is 
the halo concentration and $r_{\rm v}$ is the virial radius) for our 
calculations.  This gives us the difference in the power law indices for the 
red and blue galaxies ($\Delta \alpha$),
\begin{equation}
\Delta \alpha \equiv \alphab - \alphar = 
\frac{\log(\mu^\prime \eta^\prime)}{\log(1 + r_i c/r_{\rm v}) - 
\log(r_i c/ r_{\rm v}) }.
\label{eq:alpha}
\end{equation} 
With this relation between $\alphar$ and $\alphab$ in hand, we can perform a 
simple search over values of $\alphar$ to find the sub-profiles that combine 
to closest match an overall profile with a given value of $\alpha$.  Since we 
know that $\Delta \alpha$ must be positive, this relation guarantees a flatter 
distribution of blue galaxies in the center of halos relative to red galaxies
and Equation~\ref{eq:rho_prime} produces relatively more blue galaxies in the 
outer regions.

With this formalism in place, the only elements missing are actual values for
the parameters in Equations~\ref{eq:numgal_red}, \ref{eq:numgal_blue}, 
\ref{eq:secgal}, and \ref{eq:eta_prime}.  Since we need the $\numgal$ 
relations to calibrate $\mu$, $\eta$, $M_\alpha$ and $\sigma_\alpha$, we begin 
with $\numgalr$ and $\numgalb$.

\begin{figure}[t]
\begin{center}
\epsfxsize=200pt \epsffile{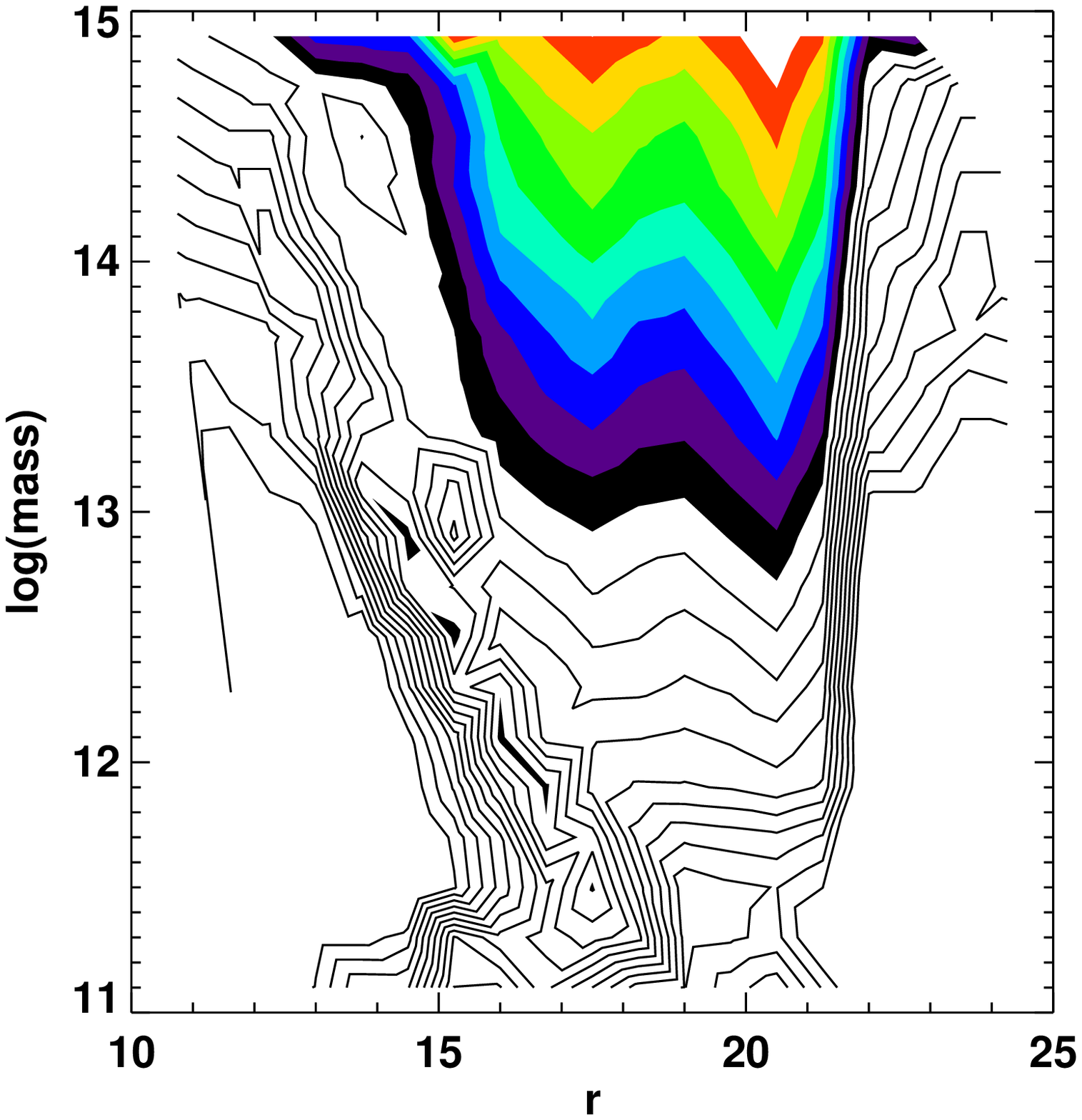} 
\epsfxsize=200pt \epsffile{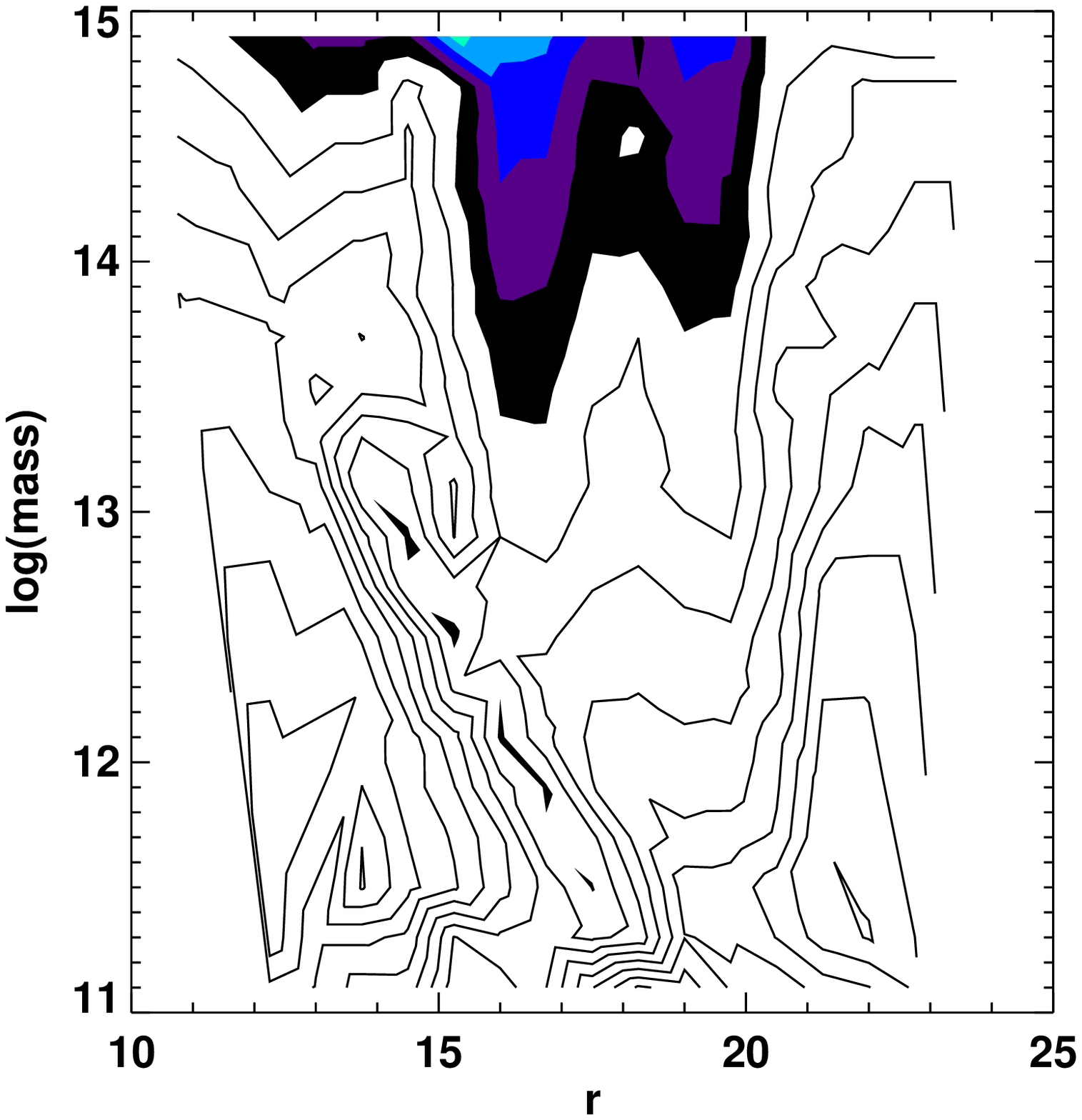}
\epsfxsize=200pt \epsffile{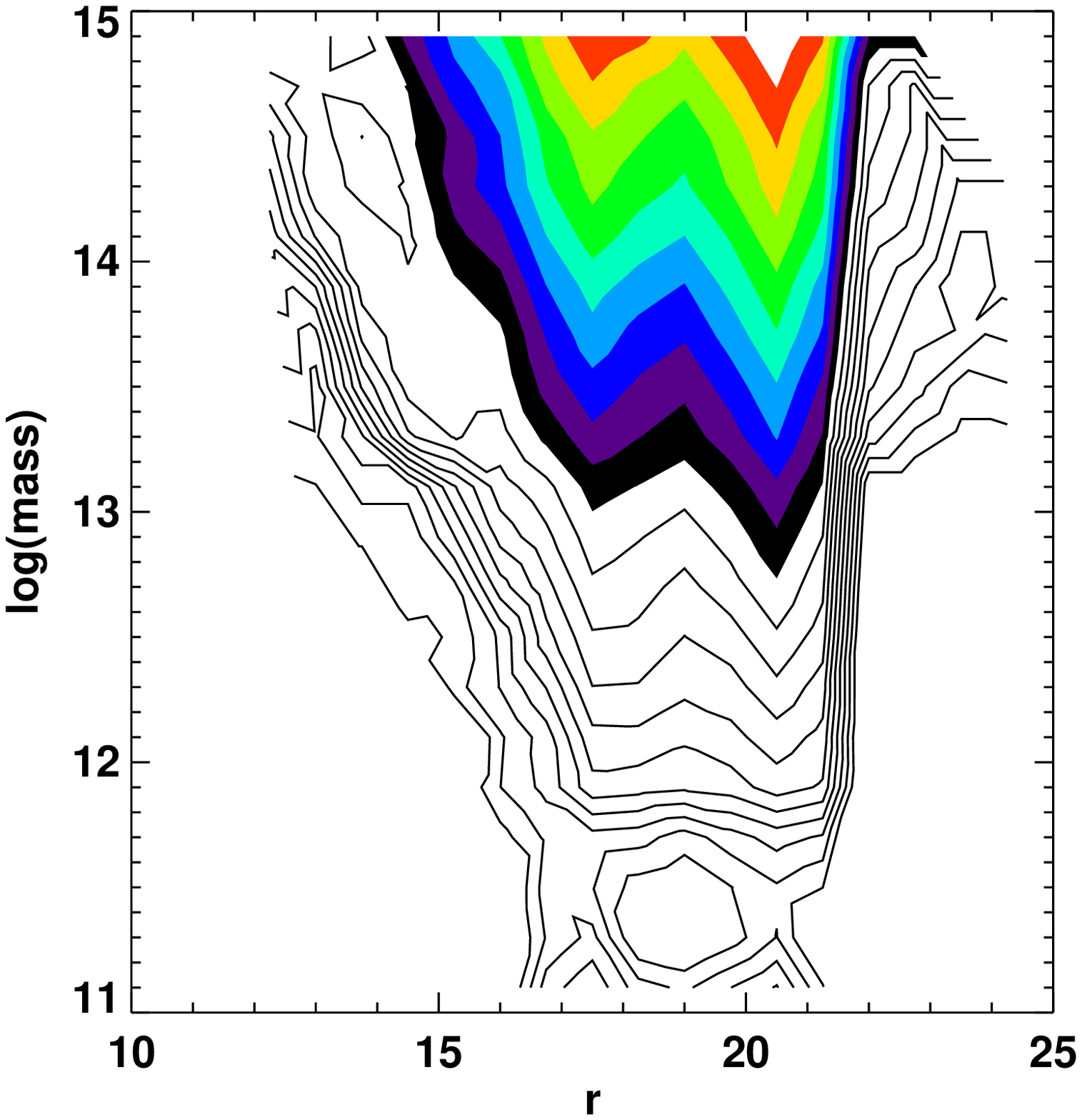} 
\caption{$\numgal$ as a function of halo virial mass and $\rprime$ magnitude
from the GIF simulations at $z = 0.06$.  From top to bottom, the panels give 
the surfaces for all galaxies, blue galaxies (rest frame 
$g^\prime - i^\prime < 0.85$) and red galaxies (rest frame 
$g^\prime - i^\prime > 0.85$), respectively.  Contours correspond to a change 
in 0.25 for $\log (\numgal)$.  Filled contours indicate $\log (\numgal) > 0$ 
and wire-frame contours indicate $\log (\numgal) < 0$.}
\label{fig:ngal_surf}
\end{center}
\end{figure}

Typically, the $\numgal$ relations are determined from simulations using very 
wide or open ended magnitude cuts.  However, as shown in 
Figure~\ref{fig:ngal_surf}, we can see that the shape of the $\numgal$ 
relations can change quite dramatically if we consider only a narrow range in 
apparent magnitude (like those described later in 
\S\ref{sec:selection_function}), particularly the location of the Gaussian 
component.  As a consequence of this complication, any constraints made on the 
parameters in $\numgalr(M)$ and $\numgalb(M)$ from a given magnitude cut will 
not accurately describe the complete number-mass relationships.  This situation 
can be salvaged to some extent however, based upon the fact that shape of the 
number-mass surface does not appear to vary strongly with redshift (the entire 
surface does shift to fainter magnitudes with increasing redshift, as one 
would expect for apparent magnitudes).  Thus, with a volume limited sample, 
one could in principle constrain the entire surface with a series of magnitude 
limited measurements.

\begin{figure}[t]
\begin{center}
\epsfverbosetrue
\epsfxsize=240pt \epsfbox{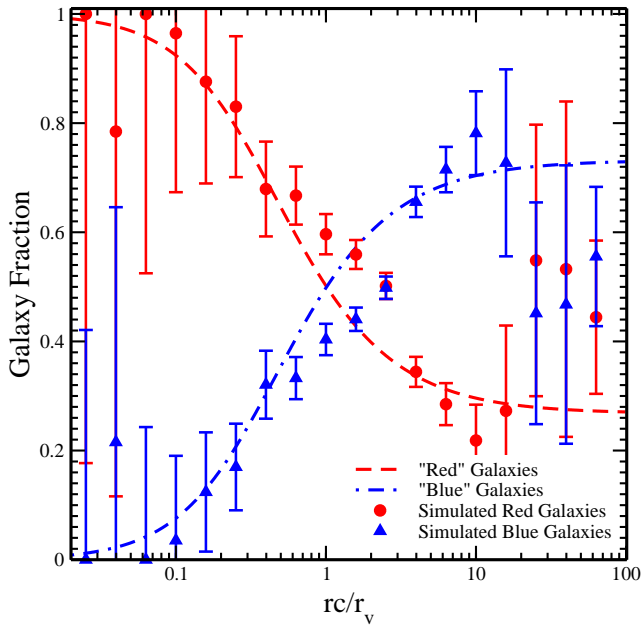}
\caption{Comparison of fraction of red and blue galaxies as a 
function of $rc /r_{\rm v}$.  The lines give the distributions predicted from 
the sub-population distributions discussed in \S\ref{sec:split} and the points
show the distribution of galaxies from the GIF simulations.  For the
simulated data, red galaxies were taken to have $\gprime - \iprime > 0.85$ 
in the rest frame (see Appendix~\ref{sec:numgal} for more details).  Error 
bars represent the Poisson error for all of the galaxies in a given radial 
bin, scaled appropriately for the fractional comparison.}
\label{fig:gal_dist}
\end{center}
\end{figure}

Leaving aside these concerns, we can fit the parameters in 
Equations~\ref{eq:numgal_red} and \ref{eq:numgal_blue} for the magnitude cut 
in \S\ref{sec:wtheta} ($20 \le \rprime \le 21$) for a typical peak redshift 
in the selection functions at $z \sim 0.3$.  The results of these fits are 
shown in Table~\ref{tab:fiducial}; a comparison of the fits to the outputs of
the simulations and a discussion of the color cut used are given in 
Appendix~\ref{sec:numgal}.  The parameters listed show a significant 
difference from previous similar fits (e.g Scoccimarro et al. (2001)).  This 
is not surprising, however, since our fits are made at higher redshift, 
resulting in a correspondingly higher mass scale for all of the $\numgal$ 
relations.  Likewise, the narrow magnitude cut leads to a relative suppression 
of the $\numgal$ relations at smaller halo mass giving larger values of 
$\gammab$ than have been reported previously.  It is interesting, however, 
that the combination of magnitude and redshift did not significantly change 
the $\secgal$ mass scale, allowing us to use similar values in 
Equation~\ref{eq:secgal} as those matching the form of 
Equation~\ref{eq:alpha_M}.  At this redshift, we also observe a higher value 
of $\mu$ in the inner regions of the halo, although the value of $\eta$ is 
largely unchanged from the $z = 0$ value in S02.  To test this set of 
parameters, we can compare the distribution of red and blue galaxies in the 
simulation to that which we would predict from our model.  As we can see in 
Figure~\ref{fig:gal_dist}, this combination of parameters reconstruct the 
halo galaxy distribution observed in the simulations reasonably well.  

\begin{figure}[b]
\begin{center}
\epsfxsize=240pt \epsffile{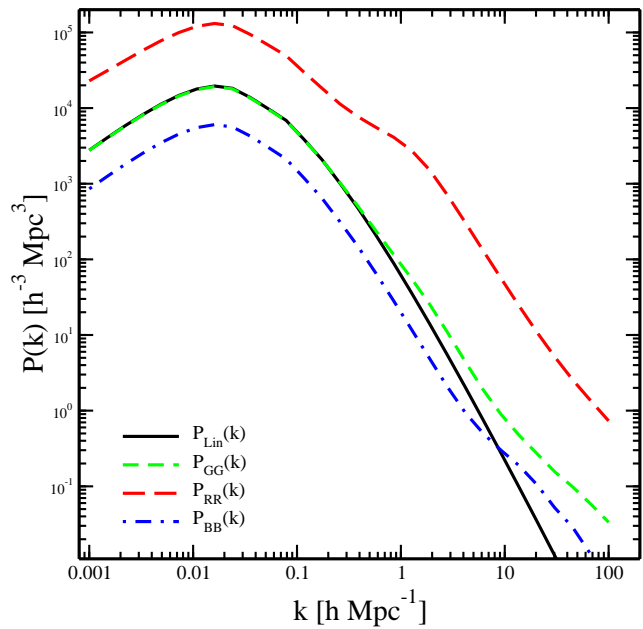}
\caption{Power spectra at $z = 0.3$ for red ($\Prr$) and blue ($\Pbb$) 
galaxies compared to the linear dark matter ($P_{\rm Lin}(k)$) and galaxy 
($\Pgg$) power spectra. \label{fig:power_spectra}}
\end{center}
\end{figure}

Putting the parameters from Table~\ref{tab:fiducial} into our prescription 
for calculating the power spectra leads to the curves shown in 
Figure~\ref{fig:power_spectra} for $z = 0.3$.  As with the $z = 0$ power 
spectra in S02, there is significant biasing in the red galaxy sample and 
anti-biasing in the blue galaxies.  Likewise, there is a similar (but weaker) 
large $k$ break in the blue power spectrum.  The power spectrum for the whole 
galaxy sample shows a slight break around $k \sim 10 h {\rm Mpc}^{-1}$ which 
is not seen in the $z = 0$ case or in low redshift observations 
(cf. Hamilton et al., 2000).  At this small scale, the power spectra are 
dominated by the smallest mass halos which, in turn, are almost exclusively 
populated by blue galaxies, particularly at higher redshifts.  This suggests 
that we should not be surprised to find a break in the power law behavior of 
$\wxx$ for a photometric redshift-selected galaxy sample.

\section{Redshift Distribution}\label{sec:selection_function}

\begin{figure}[t]
\epsfxsize=240pt \epsffile{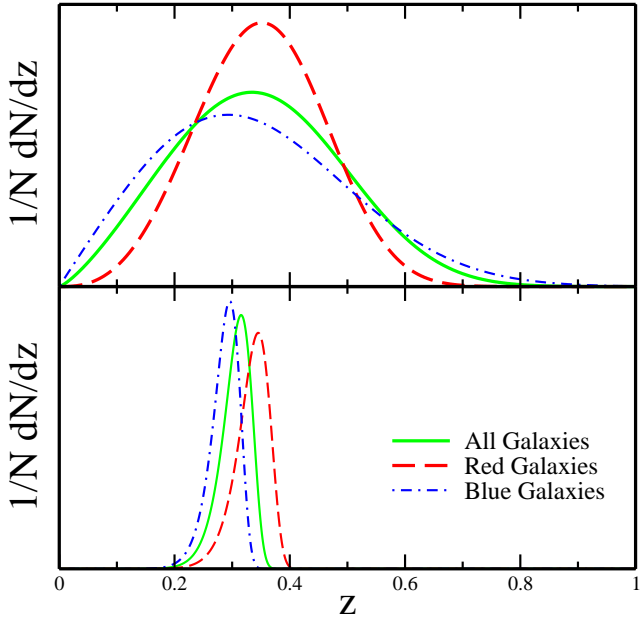} 
\caption{The top panel gives the normalized redshift distributions for all 
galaxies with $20 \le r^\prime \le 21$ and the red and blue sub-populations, 
as modified for SDSS colors from the CNOC2 survey.  The bottom panel gives an 
example of narrow redshift distributions possible for a sample of galaxies 
using photometric redshifts.}
\label{fig:dndz}
\end{figure}

To project the three dimensional power spectrum onto the sky we need to 
know the redshift distributions of the red and blue galaxies in our sample.
In order to separate the galaxies into red and blue classes, we would need
to have photometric redshift information available on each galaxy, meaning
that we could, in principle, choose the galaxy sample to match any selection
function we desired within the bounds of the survey limits.  Alternatively,
we could use the redshift distribution resulting from a simple magnitude cut.
The magnitude cut results in a somewhat broad distribution of redshifts, so 
we will use the arbitrary redshift distribution to explore the effects of a 
much narrower redshift window.  In either case, however, we are limited by the 
ability of the photometric redshift calculations to cleanly separate our galaxy
sample into red and blue types, which, in turn, is limited by the photometric
errors at a given magnitude.  Based on early results from the photometric 
redshift work on the SDSS, the practical limit for reliable redshifts is 
$\rprime \approx 21$.  

\begin{table}[t]
\begin{center}
\caption{Redshift distribution parameters for a simple magnitude cut and 
narrow window distribution.}
\label{tab:selection}
\begin{tabular}{@{}ll|c|ccc}
& Distribution & Galaxy Type & $a$ & $z_0$ & $b$ \\
\hline
& Magnitude Cut & Early & 2.6 & 0.37 & 3.3  \\  
&  & Intermediate & 2.0 & 0.37 & 2.45  \\  
&  & Late  & 0.98 & 0.4 & 2.28  \\  
&  & All  & 1.28 & 0.42 & 2.94  \\ \hline
& Narrow Window & Early & 12 & 0.35 & 15  \\  
&  & Intermediate & 12 & 0.3 & 15  \\  
&  & Late & 12 & 0.3 & 15  \\  
&  & All & 12 & 0.32 & 15  \\  

\hline 
\end{tabular}
\end{center}
\end{table}

For the case of a simple magnitude cut, we can follow the method used by 
Dodelson et al (2001) in modifying the redshift distributions found from the 
CNOC2 survey (Lin et al, 2000) to match the SDSS filters.  The redshift 
distributions for red and blue galaxies (taken from Lin's early type and the
combination of intermediate and late types, respectively) can be seen in 
the left panel of Figure~\ref{fig:dndz}.  Since these distribution functions
are taken from the morphological types rather than the color cut mentioned
in Table~\ref{tab:fiducial} and Appendix~\ref{sec:numgal}, we do not expect 
these distributions to exactly match those found in the final SDSS data, but
they should give us a reasonable approximation.  The shape of these 
distributions can be well approximated by a function of the form
\begin{equation}
\frac{d\Nx}{dz} \sim z^a \exp\left (-(z/z_0)^b \right).
\end{equation} 
The fits to the parameters in this function for red and blue galaxies with 
$20 \le \rprime \le 21$ are given in Table~\ref{tab:selection}.  We also can 
use this form to specify the shape of the narrow redshift distribution,
adopting the second set of parameters in Table~\ref{tab:selection}, resulting
in the redshift distributions shown in the right panel of 
Figure~\ref{fig:dndz}.  By design, these artificial distributions have roughly 
the same peak in redshift as the magnitude cut distributions, which should 
make the eventual comparison independent of evolution effects in the power 
spectrum.  In addition, they all avoid the region around $z \sim 0.4$ as much 
as possible.  This is particularly important due to the limitations of the 
photometric redshifts available in with the SDSS filters; there is a 
degeneracy in the colors for early type galaxies near a redshift of 0.4 
resulting in large uncertainty in the photometric redshift for these objects.

\section{Calculating $w(\theta)$ and Covariance Matrices}\label{sec:wtheta}

\subsection{Limber's Equation}

With the power spectra and selection functions in hand, we can calculate the
expected angular correlations for the red and blue galaxies ($\wrr$ and 
$\wbb$, respectively) using Limber's equation:
\begin{equation}
\wxx = \frac{1}{2 \pi} \int dk k \int d\chi \Fx \Pxxz J_0(k \theta \chi) 
\label{eq:first_limber}
\end{equation}
where $\chi$ is the comoving angular diameter distance and $J_0$ is the Bessel 
function.  We could also consider the angular cross-correlation between red
and blue galaxies, but since we will be re-casting the angular correlations in
terms of relative biases in \S\ref{sec:bias}, the cross-correlation will not 
yield any additional information.  The normalized redshift distributions, 
$\Fr$ and $\Fb$, are given by 
\begin{equation}
\Fx = \left [ \frac{1}{\Nx}\frac{d\Nx}{dz}\frac{dz}{d\chi} \right ]^2, 
\end{equation}
where
\begin{equation}
\frac{dz}{d\chi} = H_0 \left [ \Omega_M(1+z)^3 + 
\Omega_K(1+z)^2 + \Omega_\Lambda \right ]^{1/2}
\end{equation}
for a given Hubble constant ($H_0$), matter density ($\Omega_M$), cosmological
constant ($\Omega_\Lambda$) and curvature ($\Omega_K = 1 - \Omega_M - 
\Omega_\Lambda$).  As mentioned previously, we will use a flat $\Lambda$CDM 
model ($\Omega_M = 0.3$, $\Omega_\Lambda = 0.7, h = 0.7$) and normalize the 
linear power spectrum such that $\sigma_8 = 0.9$ and $n_S = 1$.  

\begin{figure}[b]
\begin{center}
\epsfxsize=240pt \epsffile{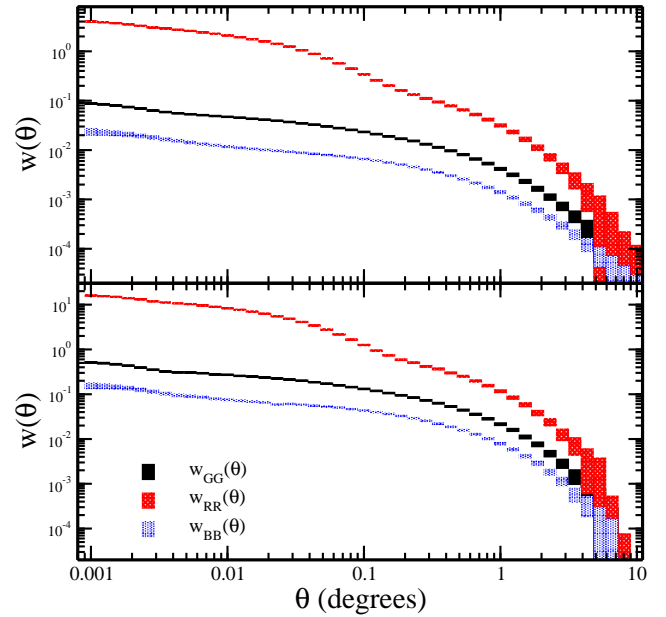}
\caption{Angular correlations for all galaxies ($\wgg$), red galaxies 
($\wrr$) and blue galaxies ($\wbb$).  The upper panel shows the angular 
correlations for the magnitude-based selection function and the lower for the 
photometric redshift selection function.}
\label{fig:correlations}
\end{center}
\end{figure}

We can simplify Equation~\ref{eq:first_limber} by assuming a linear scaling 
relation for the power spectrum, $\Pxxz = D^2(\chi) \Pxx$, where we choose 
$\Pxx$ to have the shape of the power spectrum at the peak redshift of $\Fx$.  
Making this substitution gives us
\begin{equation}
\wxx = 4 \pi^2 \int dk k \Pxx \int d\chi \Fx D^2(\chi) J_0(k \theta \chi). 
\label{eq:limber}
\end{equation}
The results of performing this calculation for the three power spectra and the 
two sets of selection function are shown in Figure~\ref{fig:correlations}.
In addition to the assumption of linear power spectrum scaling, we also 
truncate our integral over wavenumber at $k = 100 h {\rm Mpc}^{-1}$; this 
does not change the calculated values of $\wxx$ due to the flat shape of the 
kernel at small $k \theta$ and allows us to avoid any complication of the 
power spectrum shape due to baryon concentration in the innermost regions of
the halo.

\begin{figure}[b]
\begin{center}
\epsfxsize=240pt \epsffile{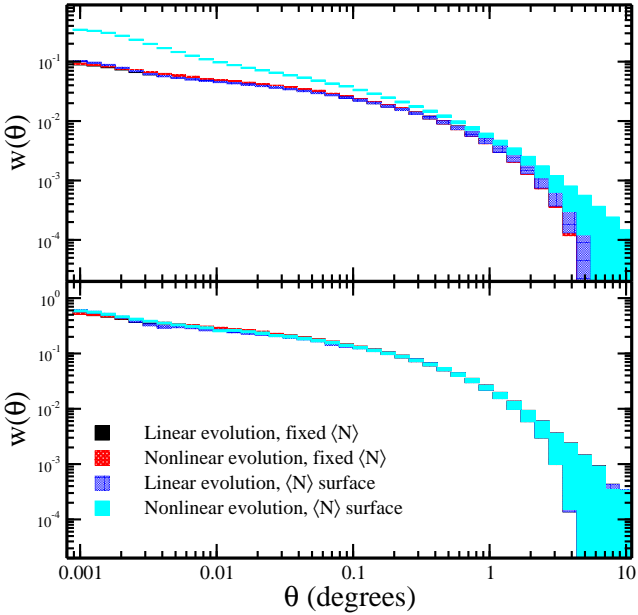}
\caption{Angular correlations for all galaxies ($\wgg$) under a number of 
assumptions for power spectrum evolution.  The black and red curves show the 
angular correlations (and associated errors) which have been calculated 
using the $\numgalr$, $\numgalb$ and $\alpha_M$ relations given by 
Equations~\ref{eq:numgal_red}, \ref{eq:numgal_blue} and \ref{eq:secgal} 
(respectively) and the parameter values in Table~\ref{tab:fiducial}.  The
black curve assumes the linear power spectrum evolution in 
Equation~\ref{eq:limber}, while the red curve calculates $\wgg$ using 
Equation~\ref{eq:first_limber}.  The blue and cyan curves take their 
$\numgal$ and $\secgal$ relations taken directly from surfaces like those 
shown in Figure~\ref{fig:ngal_surf} (shifted in magnitude appropriately for a 
given redshift) to calculate the power spectrum under linear and non-linear 
assumptions of Equations~\ref{eq:limber} and \ref{eq:first_limber}, 
respectively.  The upper panel shows the angular correlations for the 
magnitude-based selection function and the lower for the 
photometric redshift selection function.}
\label{fig:limber}
\end{center}
\end{figure}

The price we pay for the assumption of linear power spectrum scaling is the 
neglect of any change in the shape of the power spectrum at large $k$ as a 
function of redshift due to nonlinear effects.  For the photometric redshift 
selection functions, this is not a serious problem, but the width of the 
magnitude-based selection function gives one pause, particularly when 
considering the very small expected errors (\S\ref{sec:covariance}).  However, 
given the nature of the surfaces in Figure~\ref{fig:ngal_surf}, it is clear 
that any changes in the shape of the power spectrum we might see will be a 
reflection of the changing $\numgal$ relations as a function of redshift.  
This makes our choice of fiducial model for $\numgalr$ and $\numgalb$ somewhat
nebulous.  We can see the effects of calculating $\wxx$ with and without the
linear approximation for the two selection functions in 
Figure~\ref{fig:limber}.  As expected, the photometric redshift selection
function shows little to no effect even when we calculate the power spectrum
using the full $\numgal$ and $\secgal$ surfaces.  For the magnitude selection
selection function, the results are identical when we hold the $\numgal$ 
relations the same, but show strong deviations if we use the $\numgal$ and
$\secgal$ surfaces to calculate $\wxx$ without the linear approximation.
Fortunately, as we will see in \S\ref{sec:results}, the equivalent  
constraints obtained with the photometric redshift selection function make
this concern moot.

\subsection{Covariance Matrices}\label{sec:covariance}

Since we will need to account for the fact that the angular bins in our 
measurement are going to be highly correlated, we must deal with the full 
covariance matrix and not just the expected errors on each bin.   For the 
expected Gaussian and non-Gaussian sample variance covariance matrices, we 
follow the prescription given in Eisenstein \& Zaldarriaga (2001) (as well as 
similar treatments in Cooray \& Hu (2001) and Scoccimarro, et al. (1999)).  
The sample variance (${\bf C}_S(\theta,\theta^\prime)$) consists of two parts, 
a component dependent only on the two-point angular correlation function (the 
Gaussian covariance) and a second piece which is a function of the four-point 
angular correlation (the non-Gaussian covariance):
\begin{eqnarray}
{\bf C}_S(\theta,\theta^\prime) &\equiv& 
\left \langle (w(\theta) - \hat{w}(\theta)) 
(w(\theta^\prime) - \hat{w}(\theta^\prime)) \right \rangle \nonumber \\
&=& {\bf C}_G(\theta,\theta^\prime) + {\bf C}_{NG}(\theta,\theta^\prime). 
\end{eqnarray}
In order to make the calculation simpler, we can re-write 
Equation~\ref{eq:limber} in terms of the angular power spectrum ($\PKxx$),
\begin{equation}
\wxx = \int \frac{K dK}{2 \pi} \PKxx J_0(K \theta),
\end{equation}
where $\PKxx$ is given in angular wavenumber space ($K \equiv k \chi$) and
\begin{equation}
\PKxx = \frac{1}{K} \int dk \Pxx \Fx = \int \frac{d\chi \Fx}{\chi^2} \Pxx.
\end{equation}
Using this formulation, we can write the Gaussian covariance matrix 
(${\bf C}_G(\theta, \theta^\prime)$) as 
\begin{equation}
{\bf C}_G(\theta, \theta^\prime) = \frac{1}{\pi A_\Omega} \int dK K 
{\cal P}^2_{\rm XX}(K) J_0(K\theta) J_0(K \theta^\prime),
\end{equation}
where the area of the survey ($A_\Omega$) is $\pi$ steradians in the case of
the SDSS.   

To calculate the non-Gaussian component, we need to generate an estimate of 
the trispectrum for our halo model.  Fortunately, as Cooray and Hu (2001) 
indicate, the majority of the non-Gaussian covariance can be accounted for by 
merely calculating the single halo contribution of the trispectrum.  This 
term, which is independent of configuration under the assumption of spherical
halos, is given by 
\begin{eqnarray}
& & \Txxxx = \\ \nonumber 
& & \frac{\rhobar}{{\bar{n}}^4} \int_{0}^{\infty} f(\nu) 
\frac{\fourgal}{M(\nu)} |y(k_1,M)| |y(k_2,M)| |y(k_3,M)| |y(k_4,M)| d\nu,
\end{eqnarray}
where $\fourgal \equiv \alpha_M^4 \numgal^4$ under the assumptions in 
Equation~\ref{eq:secgal}.  As with the case of the Gaussian component, we 
need to project the trispectrum into angular wavenumber space, 
\begin{equation}
\TKxxxx = \int d \chi \frac{F^2_{\rm XX}(\chi)}{\chi^6} \Txxxx.
\end{equation}
Finally, the angular trispectrum needs to be averaged over an annulus in 
angular wavenumber space,
\begin{equation}
\TbarKxx = \int \frac{d^2 K_1}{A_r} \int \frac{d^2 K_2}{A_r} 
{\cal T}_{\rm 4X}(K_1,K_1,K_2,K_2),
\end{equation}
where $A_r$ is the area of the annulus.  In the limit of narrow binning, we 
can approximate $\TbarKxx$ by calculating 
\begin{equation}
T_{\rm XX}(k_1,k_2) = \frac{\rhobar}{{\bar{n}}^4} \int_{0}^{\infty} f(\nu) 
\frac{\fourgal}{M(\nu)} y^2(k_1,M) y^2(k_2,M) d\nu,
\end{equation}
and appropriately transforming into angular wavenumber space.  With this 
in hand, we can calculate the non-Gaussian component of the sample variance
using 
\begin{eqnarray}
& & {\bf C}_{NG}(\theta, \theta^\prime) = \\\nonumber 
& & \frac{1}{4 \pi^2 A_\Omega} \int dK K \int dK^\prime K^\prime
\bar{{\cal T}}_{\rm XX}(K,K^\prime) J_0(K\theta) J_0(K^\prime \theta^\prime).
\end{eqnarray}

\begin{figure}[t]
\begin{center}
\epsfxsize=240pt \epsffile{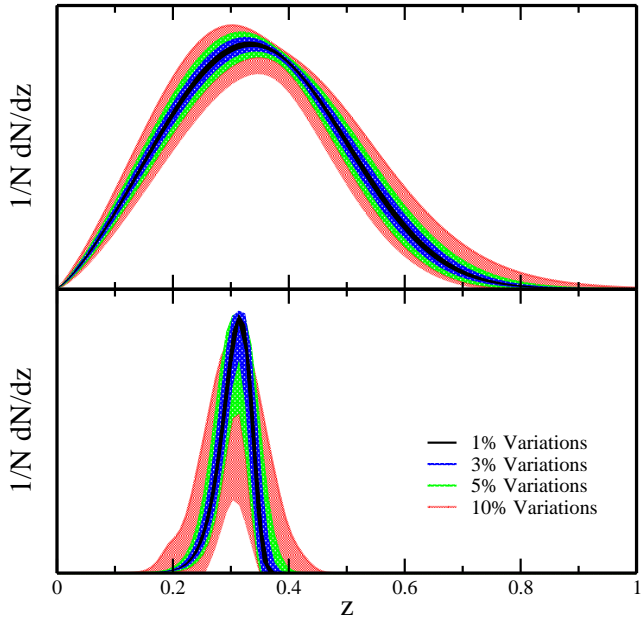}
\caption{Variations in the redshift distribution function errors.  The top
panel gives the variations for the magnitude-based selection function and
the bottom for the photometric redshift selection function.}
\label{fig:dndz_err}
\end{center}
\end{figure}

As the final component of the statistical error, we can add a Poisson term 
(${\bf C}_P(\theta,\theta^\prime)$) to the diagonal elements,
\begin{equation}
{\bf C}_P(\theta,\theta^\prime) = \frac{A_\Omega}{N^2 \delta \Omega} 
\delta_{\theta,\theta^\prime},
\end{equation}
where $N$ is the total number of galaxies, $\delta \Omega$ is the area of the
angular bin and $\delta_{\theta,\theta^\prime}$ is the standard Kronecker 
delta.  For the full photometric survey, the SDSS will contain on order 200 
million objects.  We can scale this appropriately for the 
$20 \le \rprime \le 21$ magnitude cut from \S\ref{sec:selection_function}, 
using 50 million galaxies for the full sample and half that for each of the
sub-samples.  For the photometric redshift selection function, we can combine
it with the magnitude selection function, resulting in 8 million galaxies
within the redshift and magnitude ranges. 

\begin{figure}[b]
\begin{center}
\epsfxsize=240pt \epsffile{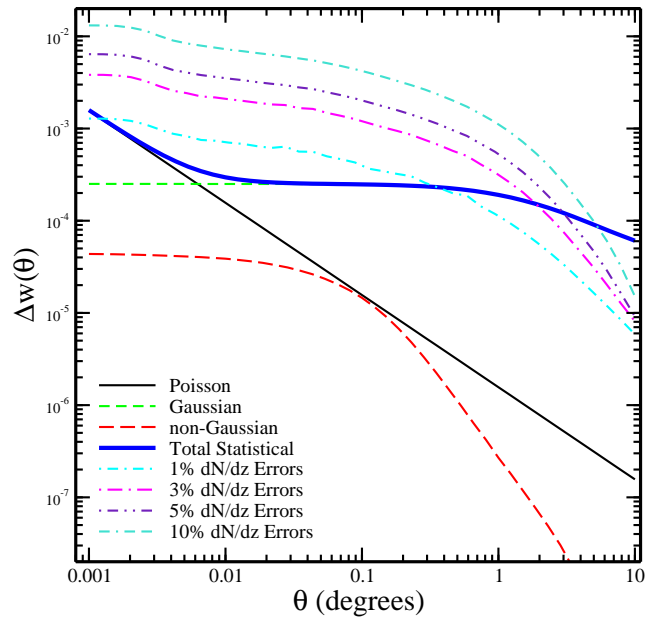}
\caption{Statistical and selection function contributions to the error on 
$\wgg$ for the magnitude-based selection function.  For the photometric
redshift selection function, the results are similar for the Poisson, Gaussian
and selection function errors.  The contribution from non-Gaussian sample 
error is relatively stronger for the photometric selection function, but 
still remains small enough to safely ignore in the total covariance matrix.}
\label{fig:covariance}
\end{center}
\end{figure}

In addition to the statistical errors due to sample variance and shot noise, 
we need to consider the errors in our calculation of the expected angular
correlations due to uncertainty in the underlying redshift distribution.  
Indeed, in \S\ref{sec:results}, we will see that, given the relatively 
large area observed and number of galaxies which will be in the final SDSS 
data set, the dominant source of error in our final constraints will come from 
the error in the redshift distributions.  To model the errors in the selection 
function, we follow the treatment given in Dodelson et al. (2001).  Since we 
do not have the exact errors for the parameter fits to the eventual redshift 
distributions, we will consider errors on those parameters of 1, 3, 5 and 10 
percent.  Drawing from 50,000 Monte Carlo realizations of the redshift 
distributions with these variations, we can calculate the covariance of the 
expected angular correlation functions (${\bf C}_Z(\theta,\theta^\prime)$) and 
use this to approximate the covariance from the redshift distribution 
uncertainty.  Figure~\ref{fig:dndz_err} shows the average deviation from the 
magnitude cut redshift distribution for each of the parameter variation 
levels.  Bringing these three pieces together gives us our final covariance 
matrix (${\bf C}(\theta, \theta^\prime)$)
\begin{equation}
{\bf C} = {\bf C}_G + {\bf C}_{NG} + {\bf C}_P + {\bf C}_{Z}.
\end{equation} 
Figure~\ref{fig:covariance} shows the error in $\wgg$ due to each term in the
covariance matrix as a function of angle.  In the absence of selection 
function errors, the errors on the very smallest angular scales are dominated 
by shot noise, giving way to Gaussian sample variance at larger angles.  
Non-Gaussian sample variance is small enough to ignore even at small angles.  
When the errors due to uncertainties in the selection function are 
included, they quickly dominate the statistical errors on small angular scales 
in all but the most modest error regime.

\section{Relative Biases}\label{sec:bias}

\begin{figure}[b]
\begin{center}
\epsfxsize=240pt \epsffile{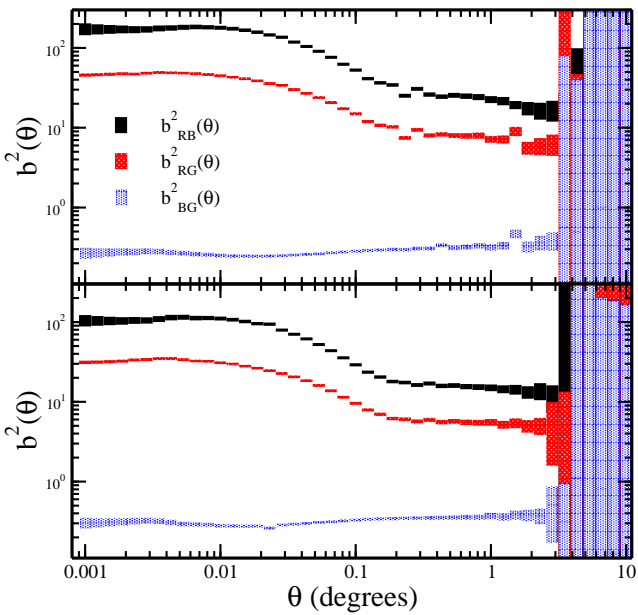}
\caption{Relative angular biases $\brb$, $\brg$ and $\bbg$ for the 
magnitude-based selection function (upper panel) and the photometric redshift 
selection function (lower panel).}
\label{fig:bias}
\end{center}
\end{figure}

In principle, the angular correlations we have calculated so far are sufficient
to calculate the parameter constraints.  However, the range of values that 
$\wxx$ takes on as a function of $\theta$ can be a problem for numerical 
derivatives.  To avoid that, we can instead use the angular biases given by 
the ratios of the angular correlations,
\begin{eqnarray}
\brb &=& \frac{\wrr}{\wbb} \nonumber \\
\brg &=& \frac{\wrr}{\wgg} \\
\bbg &=& \frac{\wbb}{\wgg} \nonumber, \label{eq:bias}
\end{eqnarray}
where $\wgg$ is the angular correlation for all of the galaxies.  As
Figure~\ref{fig:bias} shows, this switch not only decreases the 
absolute range of we must consider, but also gives us more features in the 
curves to help determine the parameters.  The price that we pay for this 
improvement is an additional step in the calculation of the covariance matrix 
we will use later in \S\ref{sec:fisher}.  Likewise, the larger amplitude 
$\wxx$'s for the photometric redshift based selection function (and 
commensurate larger values in the covariance matrix) results in proportionally 
larger errors on the resulting relative biases.

In order to properly determine the errors on the biases, we need to take into 
consideration the correlations between the angular bins in $\wxx$ as 
indicated by each correlation function's covariance matrix.  To do this, we 
decompose the covariance matrix (${\bf C}$) into its eigenmodes, creating a 
basis (${\bf R}$) where each of the modes is independent and the 
variance on that mode is given by the associated eigenvalue ($c_i$).  We can 
project $\wxx$ into that basis as
\begin{equation}
\wxx^\prime = {\bf R}^T\wxx,
\end{equation}
where ${\bf R}^T$ is the transpose of ${\bf R}$.  Since we 
require that the covariance matrix be positive definite, we set any $c_i$ 
which is negative due to numerical errors to zero and remove that mode from 
${\bf \rprime}$.  Within this basis, we can create a set of Monte Carlo 
realizations of each $\wxx^\prime$ which, when transformed back into the 
angular basis, will have the correct covariance.  Thus, the mean ratios of 
these realizations will give us the values for Equation~\ref{eq:bias} with the 
proper correlation between angular bins and the covariance between these 
realizations gives us the correct errors on our relative biases.

\section{Fisher Matrix Calculation}\label{sec:fisher}

To estimate the expected errors on the parameters in Table~\ref{tab:fiducial},
we can use the standard Fisher matrix formalism.  Choosing a fiducial set of 
parameters leads to a reference angular bias, $\bb$.  We can approximate the 
likelihood for some variation of the parameters as 
\begin{equation}
{\cal L} = \frac{1}{\left (2\pi C^D_B) \right)^{N/2}} 
\exp\left[ -\frac{1}{2} \delta(\theta) 
{\bf C}^{-1}_B(\theta, \theta^\prime) \delta(\theta^\prime) \right],
\label{eq:likelihood}
\end{equation}
where $N$ is the number of angular bins, ${\bf C}_B(\theta,\theta^\prime)$ is 
the covariance matrix for the angular biases from the Monte Carlo calculations
in \S\ref{sec:bias}, $C^D_B$ is the determinant of 
${\bf C}_B(\theta,\theta^\prime)$ and 
$\delta(\theta) \equiv (b^2(\theta) - \bb)$.  The Fisher matrix is related to 
the likelihood function as
\begin{equation}
{\bf F}_{\alpha \beta} = -\left \langle \frac{\partial \ln {\cal L}}
{\partial x_\alpha \partial x_\beta} \right \rangle,
\label{eq:fisher_def}
\end{equation}
and $({\bf F}^{-1})_{\alpha \beta}$ gives us the covariance between parameters 
$\alpha$ and $\beta$ marginalized over all other parameters, while
$({\bf F}_{\alpha \beta})^{-1}$ gives us the covariance without marginalizing 
over all other parameters.  Plugging our likelihood function from 
Equation~\ref{eq:likelihood} into Equation~\ref{eq:fisher_def} gives us the
Fisher matrix in terms of first derivatives,
\begin{equation}
{\bf F}_{\alpha \beta} =  
\frac{\partial b^2(\theta)}{\partial x_\alpha} 
{\bf C}_B^{-1}(\theta,\theta^\prime) 
\frac{\partial b^2(\theta^\prime)}{\partial x_\beta}.
\label{eq:fisher}
\end{equation}

In principle, only two of the relative biases given in Equation~\ref{eq:bias}
would be necessary to constrain the halo model parameters.  However, given the
amplitude of the errors when the effects of errors in the selection function
are included (particularly in the 10\% case), we find a more stable solution
when using all three relative biases.  This makes our $b^2(\theta)$ a 
concatenation of $\brb$, $\brg$ and $\bbg$ and requires us to calculate a 
joint covariance matrix for all biases.  In order to eliminate the presence
of singular modes in this joint matrix, we only consider those angular bins 
from $\brb$, $\brg$ and $\bbg$ which have errors less than the amplitude of 
the bias in all three measurements, reducing our number of angular bins by 
about one third.  The addition of the third relative bias does result in some
degenerate modes in our covariance matrix, but a standard singular value 
decomposition routine can handle these modes adequately, resulting in 
numerically identical Fisher matrices for either 2 or 3 relative biases in the 
small selection function error cases. 

\begin{figure}[t]
\begin{center}
\epsfxsize=240pt \epsffile{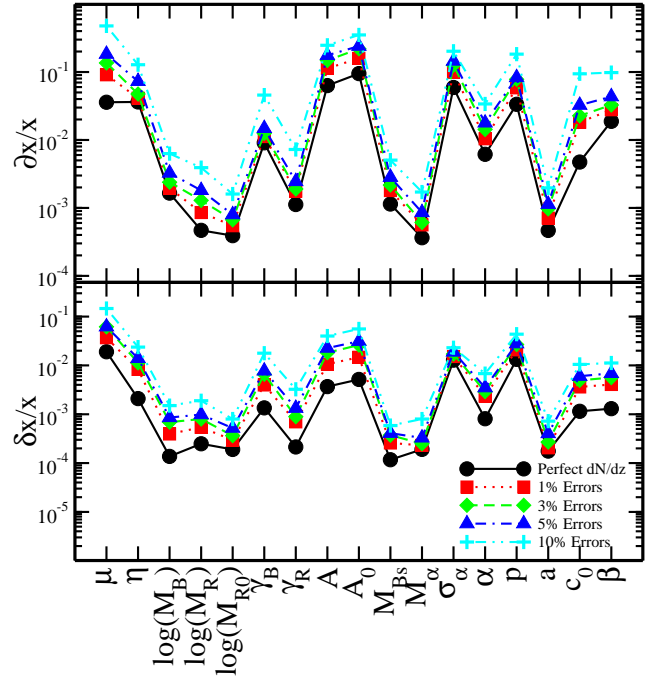}
\caption{Marginalized and unmarginalized errors ($\partial x$ and $\delta x$,
respectively) for each parameter in each of the selection function error 
regimes for the magnitude-based selection function.}
\label{fig:fractional_error_mag}
\end{center}
\end{figure}

\begin{figure}[t]
\begin{center}
\epsfxsize=240pt \epsffile{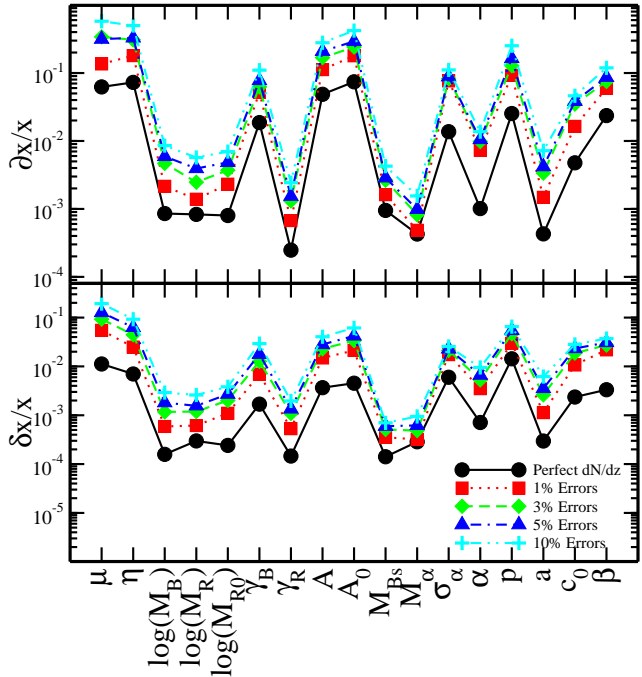}
\caption{Same as Figure~\ref{fig:fractional_error_mag}, but for the 
photometric redshift selection function.}
\label{fig:fractional_error_nar}
\end{center}
\end{figure}

In addition to these concerns, there is also the need to use a sufficient 
number of Monte Carlo realizations to ensure that the statistical noise from 
the realization (which goes as $\delta \sim 1/\sqrt{N}$ for $N$ realizations) 
is sufficiently low to allow for an accurate numerical calculation of 
the derivatives in Equation~\ref{eq:fisher}.  To meet this requirement, we 
used one million Monte Carlo realizations for each derivative calculation, 
achieving a statistical noise in the derivatives on order $0.1\%$.  Given 
this level of precision, we calculated the derivatives using centered 
derivatives with a typical step size ($\Delta x/x \equiv \delta^{1/3}$) of 
10\% for each parameter.  Given the much larger absolute value of $\Mr$, $\Mb$,
and $\Mc$ than the remainder of our parameters, we calculated the estimated 
constraint on the logarithm of each of these parameters, rather than the full 
value.  Because of this transformation we reduced the step size for these 
parameters (as well as $\Mbs$ and $M_\alpha$) to 5\%.  The accuracy of the 
resulting Fisher matrix was verified by comparison with a Fisher matrix 
calculated from the estimates of the noise in each derivative, typically 
resulting in a noise on the diagonal of the Fisher matrix of less than half a 
percent.

\section{Results}\label{sec:results}

As with any such calculation, there are essentially two questions to be 
addressed: what is the expected magnitude of the errors on each parameter
and what are the expected degeneracies between the various parameters.  We 
will take these questions in turn.

\subsection{Error Magnitudes}

Figures~\ref{fig:fractional_error_mag} and \ref{fig:fractional_error_nar}
give the fractional errors ($\partial x/x$ and $\delta x/x$) for each of the 
selection functions and selection function error regimes, where we take 
$\partial x$ to be the marginalized error on parameter $x$, and $\delta x$ to 
be the unmarginalized error:
\begin{eqnarray}
(\partial x)^2 &\equiv& ({\bf F}^{-1})_{xx} \\ \nonumber
(\delta x)^2 &\equiv& ({\bf F}_{xx})^{-1}.
\end{eqnarray}
The immediately striking aspect of each of these plots is the minuscule 
expected error in $\log(\Mb)$, $\log(\Mr)$ and $\log(\Mc)$, particularly in 
the right panel.  This is to be expected, however, since these are the 
fractional errors in logarithmic quantities.  Translated to errors on the 
actual mass scales, these correspond to roughly 1\% marginalized errors for 
both selection functions in the absence of selection function error.   
Additionally, we can see that, while the fractional errors are less 
impressive for the magnitude-based selection function, the errors using 
this selection function are more robust against errors in the selection 
function.  In general, we can see that there is not an enormous difference
in expected fractional error between the two selection functions for most 
parameters.  The magnitude-based selection function appears to be marginally
more robust against increasing selection function error.  This probably 
does not compensate for the short-comings of making the linear assumption 
shown in Figure~\ref{fig:limber}, but is worth bearing in mind if there is 
significant error in the photometric redshift (e.g. large errors due to color 
degeneracies). 

For the general halo model parameters, the expected errors are quite small.
Indeed, given the disparity between values of $\alpha$ for the NFW and Moore
profiles, as well as the scatter on the concentration parameter listed in
Bullock et al. (2001), it is possible that the expected errors on these 
parameters could be smaller than the associated errors from simulations.  For 
the sub-population parameters, the expected errors on $\eta$ and $\mu$ are 
larger than the Poisson errors on these parameters from the GIF simulations.
However, they should be sufficient to determine that they are different from 
unity, and hence the usefulness of the formalism developed in 
\S\ref{sec:split}.  The shape of the Gaussian component (as given by $A$ and 
$A_0$) in $\numgalb$ is not as well determined by the angular correlations as 
by the simulations, but the errors on the mass scales and power law indices 
($\Mr$, $\Mb$, $\Mc$, $\Mbs$, $\gammar$ and $\gammab$) should be.  Finally, 
the mass scale for the deviation from a Poisson distribution in the 
$\secgal$ relation ($M_\alpha$) should be constrained as well from the 
angular correlations as in the simulations, but the rate of that deviation
($\sigma_\alpha$) is not.  As in the case of $A$ and $A_0$, this constraint 
may improve given a different set of $\numgal$ parameters. 

\begin{figure}[b]
\begin{center}
\epsfxsize=240pt \epsffile{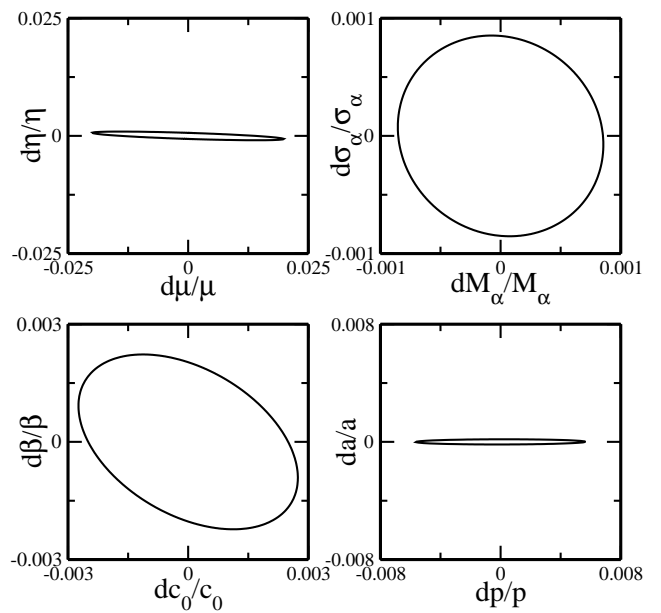}
\caption{Expected error ellipses for four combinations of parameters, 
normalized by the respective parameter values, for the magnitude-based 
selection function.  Clockwise from upper left: $\mu$ vs. $\eta$, 
$M_\alpha$ vs. $\sigma_\alpha$, $a$ vs. $p$ and $c_0$ vs. $\beta$.  Similar
results were obtained for the photometric redshift selection function, with
the exception of the $\mu$ vs $\eta$ plot, which was nearly orthogonal.}
\label{fig:error_ellipse}
\end{center}
\end{figure}

\begin{figure}[b]
\begin{center}
\epsfxsize=240pt \epsffile{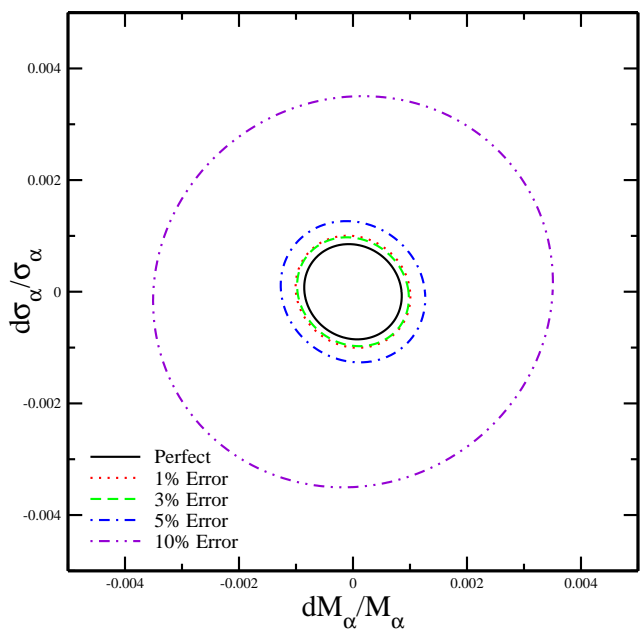}
\caption{Expected error ellipses for $M_\alpha$ \& $\sigma_\alpha$ using the 
magnitude-based selection function in each of the five selection function 
error regimes.}
\label{fig:error_ellipse_comp}
\end{center}
\end{figure}

\subsection{Parameter Degeneracies}

There are two means by which we can examine the degeneracies between the 
various halo model parameters.  First, we can look at the error ellipses 
between various parameters we expect to be correlated (e.g. $\mu$ \& $\eta$,
$c_0$ \& $\beta$, $M_\alpha$ \& $\sigma_\alpha$).  While useful for
considering particular pairs of parameters, this approach does not reveal
the full extent of the correlations between all the parameters.  To examine
this, we can decompose the Fisher matrix into its eigenvectors.  Provided
the Fisher matrix is not singular, these eigenvectors define a basis
of orthogonal combinations of parameters and tell us which combinations of 
halo parameters are naturally constrained by the angular correlations.

The error ellipse for each pair of parameters can be constructed taking the 
four corresponding elements of ${\bf F}$, inverting and decomposing the 
resulting matrix into its eigencomponents.  This effectively fixes
all of the other parameters in ${\bf F}$, resulting in errors on each 
parameter on order those found by taking $({\bf F}_{xx})^{-1/2}$.  
Figure~\ref{fig:error_ellipse} gives the expected error ellipses (normalized
by the respective parameter values) for the three combinations of parameters 
listed above, as well as the expected error ellipse for $a$ \& $p$, in the 
limit of a perfect magnitude-based selection function.  In all four cases, as 
with the other 149 combinations of parameters, the shape of the error ellipse 
is not discernibly different for the two selection functions, although the 
size will vary between selection functions according to the relative 
values of $\delta x$ for the two selection functions.  As we can see, our 
model breaks most of the degeneracy between $\mu$ and $\eta$ seen in the model 
used in S02, resulting in a nearly orthogonal error ellipse.  In contrast, we 
can see rather strong correlations between the expected errors for our 
concentration relation.

In addition to looking at correlations between parameters, we can also look
at how those correlations vary as we increase the contribution to our error
calculations from the selection function.  For the case of $M_\alpha$ \& 
$\sigma_\alpha$, we can see from Figure~\ref{fig:error_ellipse_comp} that, 
along with the expected increase in the size of the error ellipse, there is 
also some degree of wavering in the degree of correlation in the various error 
regimes.  In general, we find that this behavior is fairly consistent for all 
of the various parameter combinations, with the exception of the 10\% error 
regime.  In this case, where the covariance on the angular bias is most 
strongly influenced by the selection function error, there are a number of 
parameter combinations ($\Mr$ and $\Mb$, for instance) where the 10\% error 
ellipses were significantly rotated from the other ellipses, as much as 90 
degrees in some cases.  Likewise, the behavior of the 10\% ellipse for a given 
pair of parameters for the magnitude-based selection function appeared to be 
of little use in predicting the orientation of the same error regime in the 
photometric redshift selection function.

Having figuratively tested the waters with the error ellipses, we can move on
to the more daunting task of examining the eigenbases for the various Fisher
matrices to determine the parameter combinations that our measurements best
constrain.  Tables~\ref{tab:eigen_mag} and \ref{tab:eigen_photo} give the 
eigenbases for the two selection functions in the limit of no selection 
function error.  In both cases, we have ignored those parameters in each 
eigenmode which contribute less than 2\% to the total amplitude of the mode 
(i.e. all those parameters whose eigenmode coefficient $e_i$ was 
$|e_i| \leq 0.14$).  As one might have guessed, there are relatively few 
modes that are simply determined by one or two halo parameters.  The exceptions
to this rule occur for the modes with the best and worst constraints, where
we find strong constraints on simple combinations of a few parameters 
($\gammar$, $a$ and $\beta$) contrasted with relatively weaker constraints on 
single parameters ($c_0$, $\eta$, $A_0$ and $\mu$).  Not surprisingly, 
members of this second set tend to also have the worst fractional 
unmarginalized errors and vice versa for the elements of the first set.

Likewise, we can see some reflection of the error ellipses in the eigenbasis.
The parameter pairs $\mu$ \& $\eta$ and $a$ \& $p$ were very nearly 
uncorrelated in Figure~\ref{fig:error_ellipse} and we can see that these 
combinations of parameters do not appear in any of the eigenmodes for either 
of the Fisher matrices.  Conversely, the $M_\alpha$ \& $\sigma_\alpha$ pair 
almost invariably appear together in the eigenmodes and with the expected 
relative signs and amplitudes.  This is not a perfect guide, however, given
the correlation between $c_0$ and $\beta$ in the error ellipses and the 
absence of an eigenmode containing both parameters.

When we add in errors due to selection function uncertainty, the effects are 
similar to those seen in the error ellipses.  The eigenmodes remain mostly 
unchanged with increasing selection function error; the contribution of each 
parameter to a given eigenmode remains identical within $\sim 2\%$ of the 
total eigenmode amplitude and appears to preserve the relative signs.  
Likewise, we see an increase in the error on each eigenmode as the selection
function error increases; errors increase by a factor of 2 for each 5\% of
selection function error.  The one case of significant change in the 
eigenbasis we do see is in the best constrained eigenmodes; the degeneracy 
between $\gammar$, $a$ and $\beta$ changes for both selection functions, 
leading to an idependent mode for $a$ in the magnitude-base selection function
and independent modes for both $\gammar$ and $a$ in the photometric case.

\section{Conclusion}

As hoped, the eventual angular clustering measurements for red and blue 
galaxies should provide strong constraints ($\sim 1-10\%$) on a wide variety 
of halo model parameters.  This remains the case for moderate levels of 
uncertainty in the selection function.  In the limit of small selection
function errors, we can achieve similar constraints using both a magnitude
and photometric redshift based selection functions.  This second
approach also allows us to have much greater confidence that we are 
constraining a simple $\numgal$ relation, rather than a weighted projection 
over the $\numgal$ surfaces in Figure~\ref{fig:ngal_surf}.  In either case,
we can see that the parameters relating to the shape of the Gaussian part of
$\numgalb$ do not strongly affect the angular clustering, but the mass scales 
and power law indices in $\numgalr$ \& $\numgalb$ should be very tightly 
constrained.  In addition, we should have sufficient constraints on the 
parameters relating the halo mass function, profile, and concentration to 
determine if the values measured from simulations are consistent with 
observations.  Finally, the completed SDSS measurements should be sufficient 
to determine if the method for distributing red and blue galaxies within a 
halo given in S02 accurately reproduces the observed clustering.

In addition to the angular correlation function work presented here, there are 
a number of other measurements one might adopt to constrain halo parameters in 
a similar fashion to the methodology presented here.  In particular, 
measurements of both strong and weak lensing should provide a great deal of 
information about the structure of dark matter halos and the subsidiary 
information in rich galaxy surveys should be sufficient to probe different 
galaxy biasing by type and color on physical scales, rather than the angular 
ones presented here.  Likewise, one could readily construct alternative 
formulations of the power spectrum (weighting by star formation rate or 
bulge-to-disk ratios, for instance) which could be used in conjunction with 
spectroscopic and morphological information to inform future simulations.  
Deeper spectroscopic surveys could also attempt to find the break in the 
galaxies power spectrum shown in Figure~\ref{fig:power_spectra} as the 
fraction of galaxies shifts toward the blue at higher redshifts.  Finally, one 
could also consider the clustering of galaxy clusters as a separate test of 
the linear behavior of the halo concentration and mass function.

\acknowledgments{We would like to thank Scott Dodelson, Josh Frieman, Wayne
Hu and Stephen Kent for many useful suggestions regarding the text and scope 
of this paper.  Likewise, we would like to thank Ravi Sheth, Bhuvnesh Jain 
and Roman Scoccimarro for a number of very useful conversations.  Additional 
thanks to Guinevere Kauffmann, Volker Springel and Antonaldo Diaferio for 
assistance with the simulation data.

Support for this work was provided by the NSF through grant PHY-0079251 as 
well as by NASA through grant NAG 5-7092 and the DOE.}

\begin{table*}
\begin{center}
\caption{Fisher eigenmodes for magnitude-based selection function.}
\label{tab:eigen_mag}
\begin{tabular}{@{}r|c}
\multicolumn{1}{c}{Parameter Basis} & Expected Error \\ \hline \hline
0.32$\gammar$ + 0.84$a$ + 0.43$\beta$  & $\pm  1.13\times 10^{-4}$ \\\hline
0.55$\gammar$ - 0.53$a$ + 0.62$\beta$  & $\pm  1.85\times 10^{-4}$ \\\hline
0.77$\gammar$ - 0.63$\beta$  & $\pm  6.79\times 10^{-4}$ \\\hline
-0.37$\log(\Mb)$ - 0.19$\log(\Mc)$ + 0.53$\gammab$ - 0.28$A$ + 0.55$\Mbs$ + 
0.27$\alpha$ + 0.16$\beta$  & $\pm  8.99\times 10^{-4}$ \\\hline
0.24$\log(\Mr)$ + 0.21$\gammab$ - 0.19$A$ + 0.90$M_\alpha$ 
& $\pm 2.29\times 10^{-3}$ \\\hline
0.46$\log(\Mr)$ + 0.39$\log(\Mc)$ - 0.30$\gammab$ + 0.24$\Mbs$ + 
0.48$\sigma_\alpha$ + 0.40$\alpha$ - 0.26$a$  
& $\pm 2.88\times 10^{-3}$ \\\hline
-0.75$\log(\Mc)$ - 0.27$\gammab$ + 0.24$\Mbs$ + 0.18$M_\alpha$ + 
0.36$\sigma_\alpha$ + 0.28$p$  & $\pm 3.75\times 10^{-3}$ \\\hline
-0.67$\log(\Mr)$ - 0.18$\gammab$ + 0.23$\Mbs$ + 0.33$M_\alpha$ + 
0.27$\sigma_\alpha$ - 0.27$\alpha$ - 0.43$p$  
& $\pm 4.37\times 10^{-3}$ \\\hline
0.34$\log(\Mr)$ + 0.31$\log(\Mc)$ + 0.19$\gammab$ + 0.38$\Mbs$ - 
0.73$\alpha$ + 0.21$p$  & $\pm 4.92\times 10^{-3}$ \\\hline
-0.32$\log(\Mr)$ + 0.33$\log(\Mc)$ + 0.20$\gammab$ + 0.29$A$ + 
0.37$\sigma_\alpha$ + 0.66$p$  & $\pm 5.82\times 10^{-3}$ \\\hline
0.41$\log(\Mb)$ + 0.62$\gammab$ + 0.29$A$ - 0.33$\Mbs$ + 
0.33$\sigma_\alpha$ - 0.32$p$  & $\pm 8.21\times 10^{-3}$ \\\hline
0.48$\log(\Mb)$ + 0.23$\Mbs$ - 0.19$\sigma_\alpha$ + 0.79$c_0$  
& $\pm 1.66\times 10^{-2}$ \\\hline
-0.55$\log(\Mb)$ - 0.36$\Mbs$ + 0.31$\sigma_\alpha$ - 
0.31$\alpha$ - 0.20$p$ + 0.52$c_0$  & $\pm 2.04\times 10^{-2}$ \\\hline
-0.31$\log(\Mb)$ + 0.80$A$ + 0.26$\Mbs$ - 0.35$\sigma_\alpha$ - 0.18$p$  
& $\pm 3.37\times 10^{-2}$ \\\hline
0.97$\eta$ + 0.20$c_0$  & $\pm 0.11$ \\\hline
$A_0$  & $\pm 0.61$ \\\hline
$\mu$  & $\pm 1.82$ \\\hline
\hline 
\end{tabular}
\end{center}
\end{table*}

\begin{table*}
\begin{center}
\caption{Fisher eigenmodes for photometric redshift selection function.}
\label{tab:eigen_photo}
\begin{tabular}{@{}r|c}
\multicolumn{1}{c}{Parameter Basis} & Expected Error \\ \hline \hline
0.85$\gammar$ - 0.52$a$ & $\pm 1.44 \times 10^{-4}$ \\\hline
0.52$\gammar$ + 0.85$a$ & $\pm 2.76 \times 10^{-4}$ \\\hline
$\beta$  & $\pm 4.89 \times 10^{-4}$ \\\hline
-0.29$\log(\Mb)$ - 0.17$\log(\Mc)$ + 0.40$\gammab$ - 0.18$A$ + 0.35$\Mbs$ + 
0.20$\sigma_\alpha$ + 0.71$\alpha$  & $\pm 8.23 \times 10^{-4}$ \\\hline
-0.30$\log(\Mb)$ - 0.18$\log(\Mc)$ + 0.42$\gammab$ - 0.24$A$ + 0.41$\Mbs$ - 
0.66$\alpha$  & $\pm 1.15 \times 10^{-3}$ \\\hline
0.28$\log(\Mc)$ - 0.14$\gammab$ + 0.26$\Mbs$ - 0.21$M_\alpha$ + 
0.83$\sigma_\alpha$ - 0.17$\alpha$ - 0.15$p$  & 
$\pm 2.28 \times 10^{-3}$ \\\hline
0.16$\log(\Mr)$ + 0.25$\gammab$ - 0.29$\Mbs$ + 0.80$M_\alpha$ + 
0.37$\sigma_\alpha$  & $\pm 3.38 \times 10^{-3}$ \\\hline
-0.15$\log(\Mb)$ + 0.68$\log(\Mr)$ - 0.42$\log(\Mc)$ - 0.30$\gammab$ + 
0.23$\Mbs$ + 0.42$p$  & $\pm 3.91 \times 10^{-3}$ \\\hline
0.25$\log(\Mb)$ + 0.18$\log(\Mr)$ - 0.34$\gammab$ - 0.39$A$ + 0.46$\Mbs$ + 
0.41$M_\alpha$ - 0.17$\sigma_\alpha$ - 0.46$p$  
& $\pm 4.23 \times 10^{-3}$ \\\hline
0.56$\log(\Mb)$ - 0.19$\log(\Mr)$ + 0.19$\log(\Mc)$ + 0.16$\gammab$ - 
0.22$A$ + 0.27$\Mbs$ + 0.68$p$  & $\pm 5.19 \times 10^{-3}$ \\\hline
0.32$\log(\Mr)$ + 0.50$\log(\Mc)$ + 0.25$\gammab$ + 0.59$A$ + 0.33$\Mbs$ + 
0.18$M_\alpha$ - 0.24$\sigma_\alpha$  & $\pm 7.54 \times 10^{-3}$ \\\hline
-0.45$\log(\Mr)$ - 0.49$\log(\Mc)$ - 0.22$\gammab$ + 0.53$A$ + 0.31$\Mbs$ + 
0.29$M_\alpha$  & $\pm 8.87 \times 10^{-3}$ \\\hline
0.60$\log(\Mb)$ + 0.33$\log(\Mr)$ - 0.36$\log(\Mc)$ + 0.48$\gammab$ + 
0.18$A$ - 0.32$p$  & $\pm 1.42 \times 10^{-2}$ \\\hline
$c_0$  & $\pm 2.76 \times 10^{-2}$ \\\hline
$\eta$  & $\pm 0.16$ \\\hline
$A_0$  & $\pm 0.34$ \\\hline
$\mu$  & $\pm 3.18$ \\\hline
\hline 
\end{tabular}
\end{center}
\end{table*}

\section{References}

\def\refe {\par \hangindent=.7cm \hangafter=1 \noindent}
\def\apj { ApJ }
\def\astroph{{\tt astro-ph/}} 
\def\aap {A \& A }
\def\ajs{ ApJS }
\def\aj{AJ}
\def\prd{Phys ReV D}
\def\apjs{ ApJS }
\def\mnras { MNRAS }
\def\apjl { Ap. J. Let. }

\refe  Benson, A.J., Cole, S., Frenk, C.S., Baugh, C.M., \& Lacey, 
C.G., 2000, mnras, 311, 793
\refe  Budav{\' a}ri, T.~;., Szalay, A.~S., Connolly, A.~J., 
Csabai, I.~;., \& Dickinson, M.\ 2000, \aj, 120, 1588 
\refe  Connolly, A. et al., 2001, submitted to \apj (\astroph0107417)
\refe  Cooray, A. \& Hu, W., 2001, \apj, 554, 56
\refe  Couchman, H.M.P, Thomas, P.A, \& Pearce, F.R., 1995, \apj, 452, 797
\refe  Diaferio, A., Kauffmann, G., Colberg, J.M., \& White, 
S.D.M., 1999, \mnras, 307, 537
\refe  Dodelson, S. et al., 2001, submitted to \apj (\astroph0107421)
\refe  Eisenstein, D.J. \& Zaldarriaga, M.\ 2001, \apj, 546, 2 
\refe  Fukugita, M., Ichikawa, T., Gunn, J.E., Doi, M., Shimasaku, 
K., \& Scheider, D.P., 1996, \aj, 111,1748
\refe  Gaztanaga, E., 2001, \astroph0106379
\refe  Gunn, J. E., \& The SDSS Collaboration, 1998,\aj, 116, 3040
\refe  Kauffmann, G., Colberg, J.M., Diaferio, A. \& White, 
S.D.M., 1999, \mnras, 303, 188
\refe  Lin, H., Yee, H. K. C., Carlberg, R. G., Morris, S. L., 
Sawicki, M., Patton, D. R., Wirth, G., \& Shepherd, C. W., 1999, \apj, 518, 533
\refe  Jing, Y.P., Mo, H.J., Borner, G., 1998, \apj, 499,20
\refe  Ma, C.-P. \& Fry, J.N., 2000, \apj, 543, 503
\refe  Mo, H.J., White, S.D.M., 1996, \mnras, 282, 1096
\refe  Moore, B., Governato, F., Quinn, T., Stadel, J., \& Lake, G., 
1999, \mnras, 261, 827 
\refe  Navarro, J., Frenk, C., \& White, S.D.M., 1996, \apj, 462, 563
\refe  Navarro, J., Frenk, C., \& White, S.D.M., 1997, \apj, 490, 493
\refe  Peacock, J.A. \& Smith, R.E., 2000, \mnras, 318, 1144
\refe  Press, W.H. \& Schechter, P., 1974, \apj, 187, 425
\refe  Pryke, C., Halverson, N.W., Leitch, E.M., Kovac, J., Carlstrom, J.E., 
Holzapfel, W.L., \& Dragovan, M.\ 2002, \apj, 568, 46. 
\refe  Scoccimarro, R., Zaldarriaga, M., \& Hui, L., 1999, \apj, 527, 1
\refe  Scoccimarro, R., Sheth, R.K., Hui, L. \& Jain, B., 2001, \apj, 546, 20
\refe  Scranton, R., 2001, submitted to \mnras, (\astroph0108266)
\refe  Scranton, R. et al., 2002, submitted to \apj (\astroph0107416)
\refe  Seljak, U., 2000, \mnras, 318, 203 
\refe  Sheth, R.~K.~\& Jain, B.\ 1997, \mnras, 285, 231 
\refe  Sheth, R., Hui, L., Diaferio, A., Scoccimarro, R., 2001, \mnras, 325, 
1288
\refe  Sheth, R. \& Tormen, G., 1999, \mnras, 308, 119
\refe  Tegmark, M., Hamilton, A.J.S., Xu, Y. 2001, submitted to \mnras, 
(\astroph0111575)
\refe  White, M., Hernquist, L., Springel, V., 2001, \apj, 550, 129
\refe  White, S.D.M \& Rees, M.J., 1978, \mnras, 183, 341
\refe  York, D. G., \& The SDSS Collaboration 2000, \aj 120, 157

\pagebreak

\clearpage 

\appendix

\section{Halo Model Power Spectra} \label{sec:power}

Rewriting Equation~\ref{eq:NFW} in terms of the concentration and the mass, we 
get
\begin{equation}
\rho(r,M) = \frac{\rhos}{\left ( rc/r_{\rm v} \right)^{-\alpha} \left (1 + 
rc/r_{\rm v} \right)^{3+\alpha} },
\end{equation}
where 
\begin{eqnarray}
r_{\rm v}^3 &=& \frac{3M}{4 \pi \Delta_{\rm V} \rhobar}, \\
\rhos &=& \frac{\Delta_{\rm V} \rhobar c^3(M)}{3} 
\left [\int_0^{c(M)} d\chi \frac{\chi^{2+\alpha}}{(1+\chi)^{3+\alpha}} 
\right]^{-1},
\end{eqnarray}
$\bar{\rho}$ is the mean matter density and $M$ is the mass of the halo.  
Since we will be working in wavenumber space when we generate the power 
spectrum, we actually need to consider the Fourier transform of the halo 
profile,
\begin{equation}
y(k,M) = \frac{1}{M} \int_{0}^{r_{\rm v}} 4 \pi r^2 \rho(r,M) 
\frac{\sin (kr)}{kr} dr, 
\label{eq:rho_tilde}
\end{equation}
where we have normalized over mass so that $y(0,M) = 1$ and $y(k > 0, M) < 1$.
Note that this implies that $\rho(r > r_{\rm v}) = 0$, truncating the mass
integration at the virial radius.  This condition can be relaxed, provided that
one scales the halo mass appropriately.

With this in hand, we can move on to the next component of the halo model,
the halo mass function, ($dn/dM$).  The form of the mass function is given in 
Equation~\ref{eq:dndM}, but we need to properly normalize it by requiring that 
\begin{equation}
\frac{1}{\rhobar} \int_{0}^{\infty} \frac{dn}{dM} M dM = \int f(\nu) d\nu = 1,
\label{eq:st_norm}
\end{equation}
for the dark matter distribution.  On nonlinear scales, we expect the halos to 
cluster more strongly than the mass, and vice versa for linear scales 
(Mo \& White, 1996).  This means we need to positively bias the clustering of 
the high mass halos relative to the low mass halos.  We can generate this sort 
of halo biasing scheme for the ST mass function using
\begin{equation}
b(\nu) = 1 + \frac{\nuprime - 1}{\delta_c} + 
\frac{2p}{\delta_c(1 + {\nuprime}^p)}.
\label{eq:halo_bias}
\end{equation}
In order for the eventual power spectrum to reduce to a linear power spectrum
on large scales, we need to impose the further constraint that 
\begin{equation}
\int_{0}^{\infty} f(\nu) b(\nu) d\nu = 1,
\label{eq:bias-norm}
\end{equation}
requiring that the biased halos with mass greater than $M_*$ be balanced out
by anti-biased halos with mass less than $M_*$.  This integral is satisfied 
automatically if we use Equation~\ref{eq:halo_bias} and have properly 
normalized $f(\nu)$.

Using just these three components, we can generate the power spectrum for the
dark matter.  However, in order to predict the galaxy power spectrum, we need
to know how many galaxies are in a given halo (under the assumption that 
the distribution of galaxies in the halo follows the halo profile).  These
$\numgal$ relations are given in Equations~\ref{eq:numgal_red} and 
\ref{eq:numgal_blue}.  The inclusion of galaxies does change the normalization 
of Equation~\ref{eq:st_norm} to
\begin{equation}
\int_{0}^{\infty} \frac{\numgal}{M(\nu)} f(\nu) d\nu = \frac{\nbar}{\rhobar},
\label{eq:f-norm}
\end{equation}
where $\nbar$ is the mean number of galaxies and $\rhobar$ is the mean matter
density at redshift $z$.

On large scales, the power spectrum is dominated by correlations between 
galaxies in separate halos.  We need to convolve the halo profile with the
mass function to account for the fact that halos are not point-like objects.
Since we are in Fourier space, we can perform the convolution using simple
multiplication.  The halo-halo power ($\Phhgg$) is then simply,
\begin{equation}
\Phhgg = \Plin \left [ \frac{\rhobar}{\nbar} \int_{0}^{\infty} f(\nu) 
\frac{\numgal}{M(\nu)} b(\nu) y(k,M) d\nu \right ]^2,
\label{eq:halo-halo}
\end{equation}
where $\Plin$ is the linear dark matter power spectrum,
\begin{equation} 
\left \langle \delta({\rm k}) \delta({\rm k}^\prime) \right \rangle  = 
(2 \pi)^3 \delta \left ({\rm k} - {\rm k}^\prime \right ) \Plin.
\end{equation}

For small scales, the dominant contribution to the power spectrum comes from 
correlations between galaxies within the same halo.  This single halo term is 
independent of $k$ at larger scales, giving it a Poisson-like behavior.  In 
order to account for the fact that a single galaxy within a halo does not
correlate with itself, we use the second moment of the galaxy number relation,
$\secgal$, to calculate the Poisson power ($\PPgg$),
\begin{equation}
\PPgg = \frac{\rhobar}{\nbar^2} \int_{0}^{\infty} f(\nu) 
\frac{\secgal}{M} |y(k,M)|^\zeta d\nu.
\label{eq:poisson}
\end{equation}
Seljak (2000) takes $\zeta = 2$ for $\secgal > 1$ and $\zeta = 1$ for 
$\secgal < 1$; this is done to account for the galaxy at the center of the 
halo in the limit of small number of galaxies.  Adding $\Phhgg$ and $\PPgg$, 
we recover the galaxy power spectrum at all wavenumbers, $\Pgg$.

\subsection{Calculating Subpopulation Power Spectra}

With the modifications to the mass distributions which go into red and blue
galaxies given by Equations~\ref{eq:rho_prime} and \ref{eq:alpha}, we need to
regenerate $y(k,M)$ for each profile.  We also need to re-normalize $f(\nu)$ 
for each sub-population using Equation~\ref{eq:f-norm} to account for the 
differences in $\numgal$ and $\nbar$:
\begin{eqnarray}
\int_{0}^{\infty} \frac{\numgalr}{M(\nu)} 
\fr(\nu)d\nu &=& \frac{\nbarr}{\rhobar} \\ 
\int_{0}^{\infty} \frac{\numgalb}{M(\nu)} 
\fb(\nu)d\nu &=& \frac{\nbarb}{\rhobar} \nonumber
\end{eqnarray}
Once this has been done, we can insert the above (along with the 
color-dependent halo profiles) into Equations~\ref{eq:halo-halo} and 
\ref{eq:poisson} to generate the power spectra for red and blue galaxies:
\begin{eqnarray}
\frac{\Phhrr}{\Plin} = 
\left [ \frac{\rhobar}{\nbarr} \int_{0}^{\infty} \fr(\nu) 
\frac{\numgalr}{M(\nu)} b(\nu) \yr(k,M) d\nu \right ]^2 \\
\PPrr  = {\rhobar} \int_{0}^{\infty} \fr(\nu) 
\frac{\secgalr}{\nbarr^2 M(\nu)} |\yr(k,M)|^\zeta d\nu, \nonumber \\
\frac{\Phhbb}{\Plin} = 
\left [ \frac{\rhobar}{\nbarb} \int_{0}^{\infty} \fb(\nu) 
\frac{\numgalb}{M(\nu)} b(\nu) \yb(k,M) d\nu \right ]^2 \nonumber \\
\PPbb = \rhobar  \int_{0}^{\infty} \fb(\nu) 
\frac{\secgalb}{\nbarb^2 M(\nu)} |\yb(k,M)|^\zeta d\nu. \nonumber
\end{eqnarray}
As before, we generate the total power spectra ($\Prr$ and $\Pbb$) by taking
the sum of these parts,
\begin{eqnarray}
\Prr &=& \Phhrr + \PPrr \\
\Pbb &=& \Phhbb + \PPbb. \nonumber
\end{eqnarray}

\section{Sub-Population Selection and $\numgal$ Parameters} \label{sec:numgal}

For all of the parameter values related to the red and blue galaxy HOD, we 
fit the relations in Equations~\ref{eq:numgal_red}, \ref{eq:numgal_blue}, and 
\ref{eq:secgal} to the galaxy catalogs produced by the GIF simulations.  The 
details of the semi-analytic methods applied in the simulations can be found 
in Kaufmann et al. (1999), but we will briefly discuss some of the relevant 
features here.

In the broadest strokes, semi-analytic methods like those applied in the GIF 
simulations take the outputs of an N-body cosmological simulation at a number 
of time-steps, determine where galaxies will have formed based on some 
prescription and let the galaxies evolve from that point in time until the 
present.  In the case of the GIF simulations, the N-body simulations were 
generated using {\it Hydra} (Couchman, Thomas \& Pearce, 1995), an adaptive 
particle-particle particle mesh code written as part of the VIRGO 
collaboration.  Four different cosmological models were used in the initial 
work, but, for the purposes of the calculations in this paper, we only used 
the $\Lambda$CDM outputs ($\Omega_M = 0.3$, $\Omega_\Lambda =0.7$, $h = 0.7$, 
$\sigma_8 = 0.9$) with the SDSS filters (not mentioned in Kaufmann et al.).  
These simulations are 141 $h^{-1}$ Mpc on a side and have a mass resolution 
on order $10^{11} h^{-1} M_\sun$.  

\begin{figure}[b]
\begin{center}
\epsfxsize=240pt \epsffile{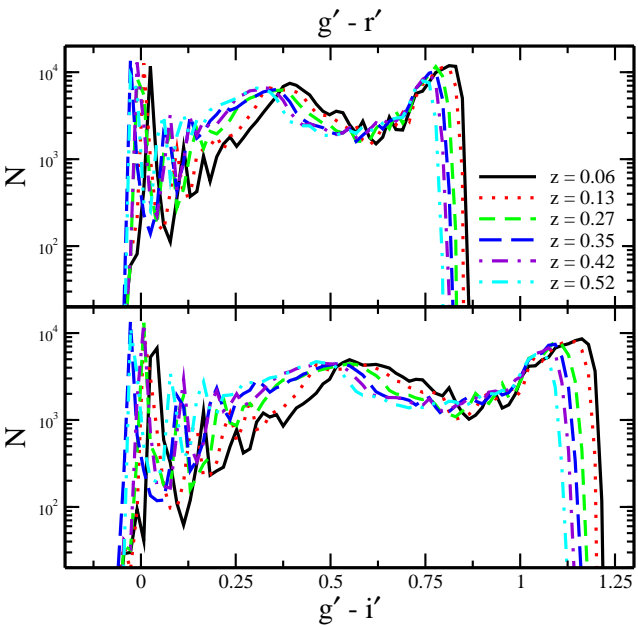}
\caption{Color distributions as a function of redshift using 
$\gprime - \rprime$ (upper) and $\gprime - \iprime$ (lower).}
\label{fig:color_dist}
\end{center}
\end{figure}

At each time step used in the GIF simulations, a friends-of-friends group 
finding routine with linking length 0.2 was applied to the N-body outputs and
each group of 10 or more particles was marked as a dark matter halo.  The 
most bound member of each such group was flagged as the central galaxy of the
halo.  In subsequent time steps, previously determined halos are checked 
against mergers with other halos.  In the case of a merger, the galaxy at the
center of the most massive progenitor halo is considered to be at the center
of the combined halo and the properties of that galaxy are transferred to the 
most bound particle of the new halo.  Galaxies associated with less massive
progenitor halos are now satellite galaxies of the new halo and remain 
associated with their original particles.  In the GIF simulation outputs used
for our calculations, each catalog contained $\sim 90,000$ halos and 
$\sim 180,000$ galaxies (of which $\sim 35,000$ fell within our apparent 
magnitude cut).

Once the positions of the individual galaxies within each halo have been 
set, the evolution of the stars in each galaxy can be determined.  Even in 
the simplest terms, this requires a number of considerations: availability of
cool gas, star formation rate, supernovae feedback, initial stellar mass 
function, metallicity (this is held at solar levels throughout the 
calculations), etc.  In addition to these intra-galactic effects, there are 
also merger effects (combination of two satellite galaxies or in-fall of 
satellite galaxies into the central halo galaxy) and the associated creation 
of galaxy bulges and star-burst activity.  All of these processes require 
tuning to one degree or another in order to reasonably reproduce observed 
luminosity functions and Tully-Fisher relations.  Since future SDSS angular 
clustering measurements discussed here will combine both the galaxy evolution 
and clustering aspects of the model (at a variety of redshifts), they should 
serve as an excellent test for many aspects of these treatments.

In splitting the GIF simulation galaxy catalogs into red and blue samples, we
had two primary considerations.  First, we wanted to produce a selection method
for the data that was robust in segregating what appear to be two rather 
distinct sub-populations.  Second, we wanted a criterion which could reasonably
be applied to actual galaxies near the limit of our magnitude selection of 
$\rprime = 21$.  The first of these requirements meant choosing a color cut
that varied between the two populations slowly enough that passive evolution in
galaxy colors over the extent of the redshift range was relatively small.
The other requirement meant restricting ourselves to the $\gprime$, $\rprime$
and $\iprime$ bands, as objects at the faint end of our magnitude cut
will often fall below the detection threshold in $u^\prime$ and $z^\prime$.

The data sets we considered for this selection consisted of 6 redshift epochs:
$z = 0.06$, $0.13$, $0.27$, $0.35$, $0.42$ and $0.52$.  
Figure~\ref{fig:color_dist} shows the distribution of rest-frame 
$\gprime - \rprime$ and $\gprime - \iprime$ colors for the galaxies at 
each of the redshift epochs.  In both cases, the distribution is roughly 
bimodal, with a spike of very blue star-forming galaxies at 
$\gprime - \rprime = \gprime - \iprime \sim 0$.  There is not an enormous
difference between the two color distributions and it is clear that a simple 
straight line cut will select a slightly different population at higher 
redshift than at lower redshift.  However, given the wider distribution of 
$\gprime - \iprime$, we should suffer from less difference with redshift
than with $\gprime - \rprime$.  With this in mind, we split our sample at 
\begin{equation}
\gprime - \iprime = 0.85.
\label{eq:color_cut}
\end{equation}

\begin{figure}[t]
\begin{center}
\epsfxsize=206pt \epsffile{Mag_Ngal_z0.06.eps} 
\epsfxsize=206pt \epsffile{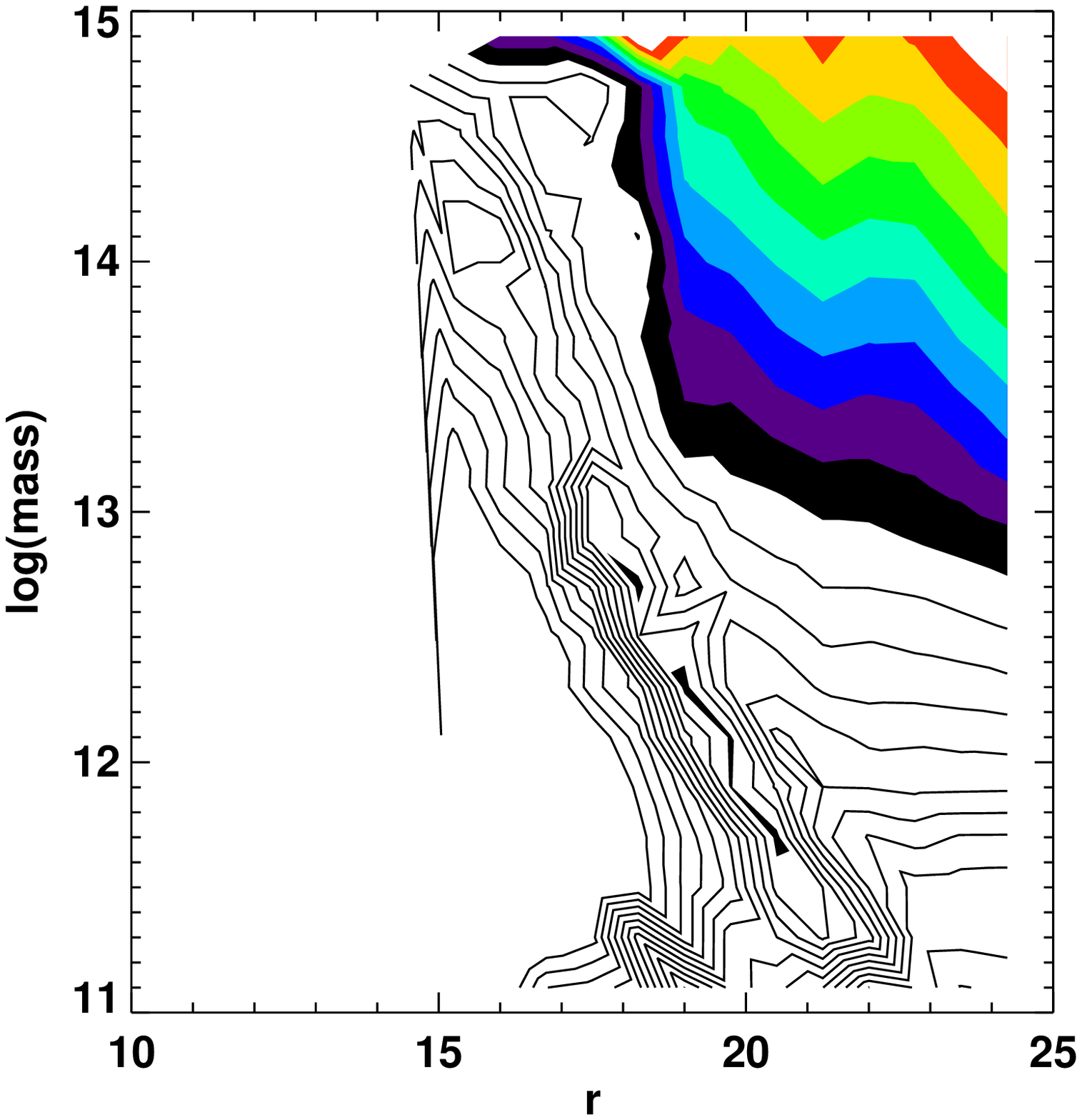} 
\epsfxsize=206pt \epsffile{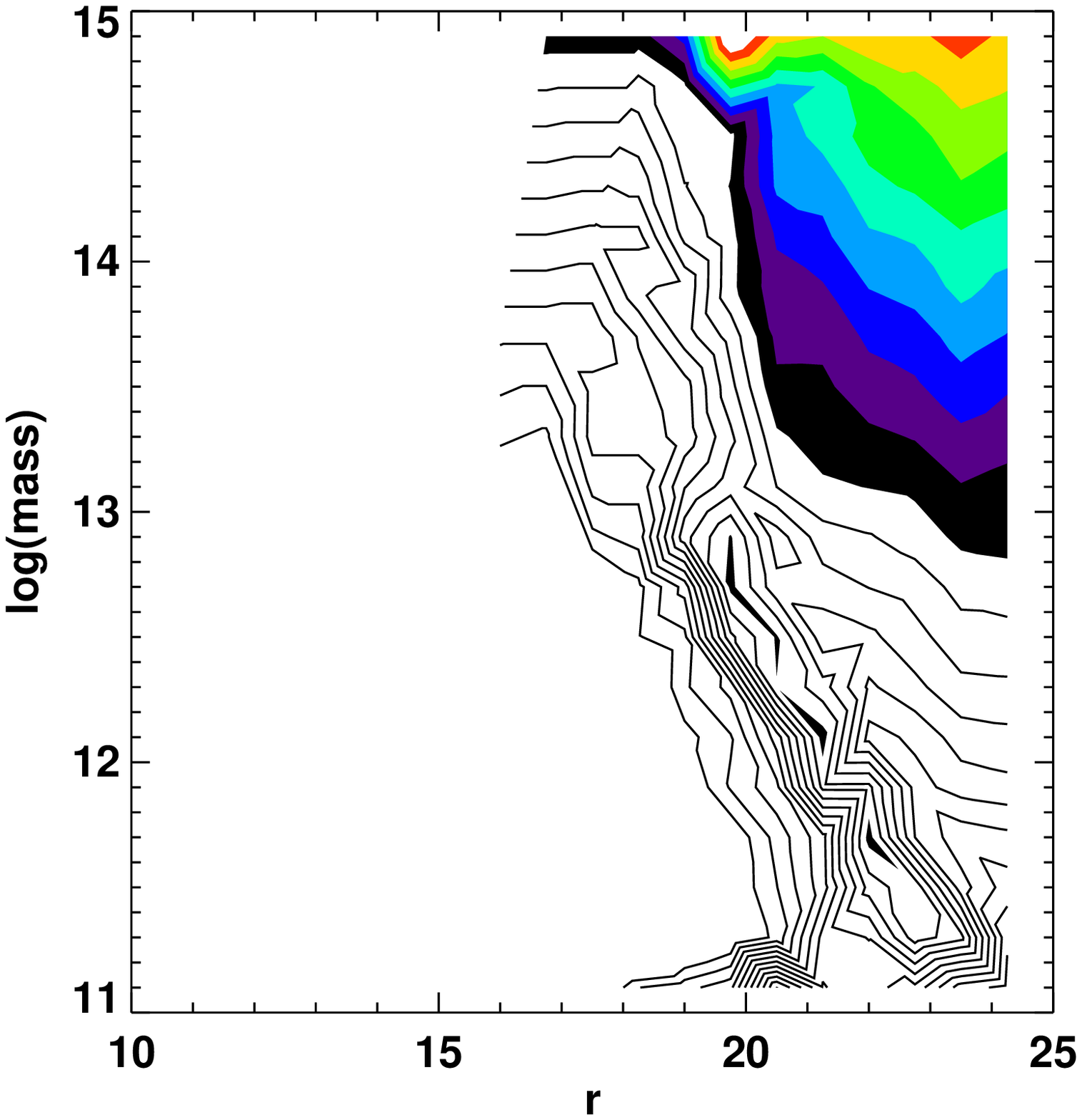} 
\caption{$\numgal$ as a function of halo virial mass and $\rprime$ magnitude
from the GIF simulations at $z = 0.06$, $0.27$ and $0.52$ (top to bottom).
As in Figure~\ref{fig:ngal_surf}, filled contours indicate $\log(\numgal) > 0$,
wire-frame contours indicate $\log(\numgal) < 0$ and successive contours 
a change of 0.25 in $\log(\numgal)$.}
\label{fig:ngalz_surf}
\end{center}
\end{figure}

We can test the effectiveness of our color selection by looking at the 
variation of the $\numgal$ surfaces as a function of redshift.  
In Figure~\ref{fig:ngalz_surf}, we plot the surfaces for the whole galaxy 
sample.  As mentioned in \S\ref{sec:fiducial}, the overall galaxy surface 
appears reasonably static as a function of redshift and the sub-population 
surfaces behave similarly.  We can also see that the shifting in apparent 
magnitude proceeds with redshift as we would expect.  

\begin{figure}[t]
\begin{center}
\epsfxsize=240pt \epsffile{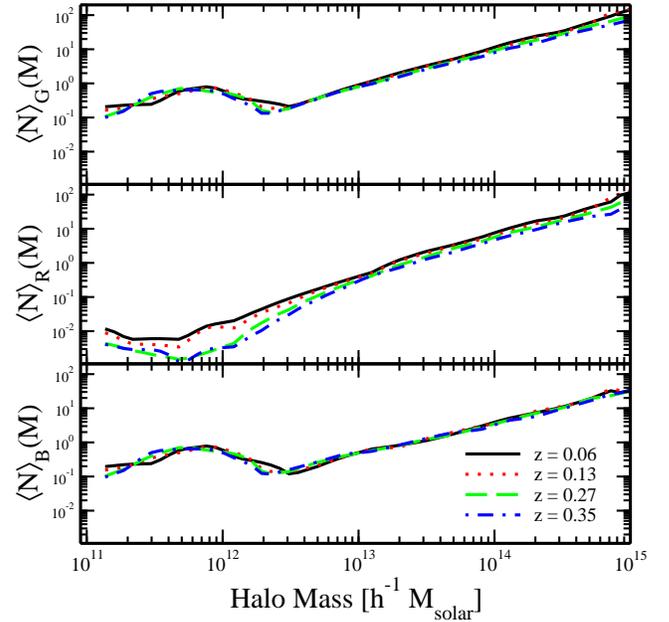}
\caption{$\numgal$ relations for all galaxies (upper), red galaxies (middle)
and blue galaxies (bottom).}
\label{fig:ngal_evol}
\end{center}
\end{figure}

For a more specific look at the possible evolution of the $\numgal$ relations
as it applies to the calculations in this paper, we can shift the surface 
from each redshift regime appropriately for the $z \sim 0.3$ selection 
function, apply our magnitude cut and compare the $\numgal$ curves.  
Figure~\ref{fig:ngal_evol} shows the $\numgal$ relations produced by this 
method for the $z = 0.06$, $0.13$, $0.27$ and $0.35$ epochs.  The $\numgalb$
curves show no significant signs of evolution.  There is some shift in the 
mass scales for $\numgalr$, both $\Mr$ and $\Mc$, but the $\gammar$ is largely 
unchanged.  In choosing our fiducial model, we fit the $\numgalr$ parameters 
from the $z = 0.27$ epoch.  While this behavior does make the modeling of 
the $\numgal$ relations more complicated than would be the case if the 
all the surfaces were static, the prediction of a red galaxy distribution 
evolving against a background of a static blue galaxy distribution is 
intriguing.

In addition to the $\numgal$ relations, we can also check the $\secgal$ 
relations for evolution.  As one might expect from Figure~\ref{fig:ngal_evol},
measurements of $\secgalr$ from the simulations were very noisy and did not
lend themselves to a reliable fit in the region where one expects strong 
deviation from a Poisson distribution.  Rather, since we chose to use a 
universal $\alpha_M$, we fit $M_\alpha$ and $\sigma_\alpha$ from 
Equation~\ref{eq:secgal} using $\secgal$ and $\numgal$ measurement of the blue 
sub-sample.  This approach may miss some of the evolution present in the red 
galaxy sample, but this seems unavoidable.  Unlike in the $\numgal$ 
measurements, more recent redshift epochs did show stronger Poisson behavior 
at lower mass scales; $M_\alpha$ stayed roughly fixed, but $\sigma_\alpha$ 
increased, leading to a more gradual decrease in $\alpha_M$ for lower mass 
halos.  As with $\numgalr$, the effect was not dramatic, but did result in 
roughly doubling $\sigma_\alpha$ from the $z = 0.06$ epoch from its value for 
the $z = 0.35$ epoch.  As before, we chose the parameter fits using the 
$z = 0.27$ surfaces for our fiducial model.  

\typeout{ ***** EDIT TEMPLATE.LOT,  TEMPLATE.LOF, & TEMPLATE.TOC! *****}

\end{document}